\documentclass{aa} 

\usepackage{graphicx}
\usepackage{txfonts}
\usepackage{natbib}
\bibpunct{(}{)}{;}{a}{}{,} % A&A style

\begin{document}

\title{The VIMOS-VLT Deep Survey\footnotemark}
\subtitle{Color bimodality and the mix of galaxy populations up to z$\sim$2}

\author{
P. Franzetti \inst{1}      % GROUP
\and M. Scodeggio \inst{1}
\and B. Garilli \inst{1}
\and D. Vergani \inst{1}
\and D. Maccagni \inst{1}
\and L. Guzzo \inst{2}
\and L. Tresse \inst{3}
\and O. Ilbert \inst{4}
\and F. Lamareille \inst{5}
\and T. Contini \inst{6}
\and O. Le F\`evre \inst{3}
\and G. Zamorani \inst{5}
\and J. Brinchmann \inst{7}
\and S. Charlot \inst{8}
\and D. Bottini \inst{1}     % BUILDERS
\and V. Le Brun \inst{3}
\and J.P. Picat \inst{6}
\and R. Scaramella \inst{9,10}
\and G. Vettolani \inst{9}
\and A. Zanichelli \inst{9}
\and C. Adami \inst{3}      % CORE
\and S. Arnouts \inst{3}
\and S. Bardelli  \inst{5}
\and M. Bolzonella  \inst{5} 
\and A. Cappi    \inst{5}
\and P. Ciliegi    \inst{5}  
\and S. Foucaud \inst{12}
\and I. Gavignaud \inst{13}
\and A. Iovino \inst{2}
\and H.J. McCracken \inst{8,14}
\and B. Marano     \inst{15}  
\and C. Marinoni \inst{16}
\and A. Mazure \inst{3}
\and B. Meneux \inst{1,2}
\and R. Merighi   \inst{5} 
\and S. Paltani \inst{17,18}
\and R. Pell\`o \inst{6}
\and A. Pollo \inst{3}
\and L. Pozzetti    \inst{6} 
\and M. Radovich \inst{11}
\and E. Zucca    \inst{5}
\and O. Cucciati \inst{2,19}
\and C.J. Walcher\inst{3}
}

%\and M. Bondi \inst{}         % ASSOCIATES  
%\and A. Bongiorno \inst{}
%\and S. de la Torre \inst{}
%\and Y. Mellier \inst{,}
%\and P. Merluzzi \inst{}
%\and S. Temporin \inst{}

\institute{
%1
INAF - IASF Milano, via Bassini 15, I-20133, Milano, Italy\\
\email{paolo@lambrate.inaf.it}
\and
%2 
INAF - Osservatorio Astronomico di Brera,Via Brera 28, Milano, Italy
\and
%3 
Laboratoire d'Astrophysique de Marseille, UMR 6110 CNRS-Universit\'e de Provence,  BP8, 13376 Marseille Cedex 12, France
\and
%4 
Institute for Astronomy, 2680 Woodlawn Dr., University of Hawaii, Honolulu, Hawaii, 96822	
\and
%5 
INAF - Osservatorio Astronomico di Bologna, Via Ranzani,1, I-40127, Bologna, Italy
\and
%6 
Laboratoire d'Astrophysique de l'Observatoire Midi-Pyr\'en\'ees (UMR 5572), 14, avenue E. Belin, F31400 Toulouse, France
\and
%7 
Centro de Astrofisica da Universidade do Porto, Rua das Estrelas, 4150-762 Porto, Portugal 
\and
%8 
Institut d'Astrophysique de Paris, UMR 7095, 98 bis Bvd Arago, 75014 Paris, France
\and
%9 
INAF - IRA, Via Gobetti,101, I-40129, Bologna, Italy
\and
%10 
INAF - Osservatorio Astronomico di Roma, Via di Frascati 33, I-00040, Monte Porzio Catone, Italy
\and
%11 
INAF - Osservatorio Astronomico di Capodimonte, Via Moiariello 16, I-80131, Napoli, Italy
\and
%12 
School of Physics \& Astronomy, University of Nottingham, University Park, Nottingham, NG72RD, UK
\and
%13 
Astrophysical Institute Potsdam, An der Sternwarte 16, D-14482 Potsdam, Germany
\and
%14 
Observatoire de Paris, LERMA, 61 Avenue de l'Observatoire, 75014 Paris, France
\and
%15 
Universit\`a di Bologna, Dipartimento di Astronomia, Via Ranzani,1, I-40127, Bologna, Italy
\and
%16 
Centre de Physique Th\'eorique, UMR 6207 CNRS-Universit\'e de Provence, F-13288 Marseille France
\and
%17 
Integral Science Data Centre, ch. d'\'Ecogia 16, CH-1290 Versoix
\and
%18 
Geneva Observatory, ch. des Maillettes 51, CH-1290 Sauverny
\and
%19 
Universit\`a di Milano-Bicocca, Dipartimento di Fisica - Piazza delle Scienze, 3, I-20126 Milano, Italy
}

\date{Received 30 June 2006 / Accepted 11 January 2007}

\abstract
{}
{
In this paper we discuss the mix of star-forming and passive galaxies up to
z$\sim$2, based on the first epoch VIMOS-VLT Deep Survey (VVDS) data.
}
{
We compute rest-frame magnitudes and colors and analyse the color-magnitude 
relation and the color distributions. We also use the multi-band VVDS photometric 
data and spectral templates fitting to derive multi-color galaxy types.
Using our spectroscopic dataset we separate galaxies based on a 
star-formation activity indicator derived combining the equivalent width 
of the [OII] emission line and the strength of the D$_n$(4000) continuum break.
}
{
In agreement with previous works we find that the global galaxy rest-frame color
distribution follows a bimodal distribution at z$\le$1, and we establish that
this bimodality holds up to at least z=1.5. 
The details of the rest-frame color distribution depend however on redshift and 
on galaxy luminosity, with faint galaxies being bluer than the luminous 
ones over the whole redshift range covered by our data, and with galaxies
becoming bluer as redshift increases. This latter blueing trend does not depend, 
to a first approximation, on galaxy luminosity.  
The comparison between the spectral classification and the rest-frame colors shows 
that about 35-40 \% of the red objects are in fact star forming galaxies. 
Hence we conclude that the red sequence cannot be used to effectively isolate a 
sample of purely passively evolving objects within a cosmological survey. 
We show how multi-color galaxy types have a slightly higher efficiency
than rest-frame color in isolating the passive, non star-forming galaxies
within the VVDS sample. 
Connected to these results is also the finding that the color-magnitude relations 
derived for the color and for the spectroscopically selected early-type galaxies 
have remarkably similar properties, with the contaminating star-forming galaxies 
within the red sequence objects introducing no significant offset in the rest 
frame colors. 
Therefore the average color of the red objects does not appear to be a very
sensitive indicator for measuring the evolution of the early-type galaxy population.
}
{}


\authorrunning{P. Franzetti et al.}
\titlerunning{The VVDS - Color bimodality and the mix of galaxy populations up to z$\sim$2}

\maketitle 

\footnotetext{* \footnotesize{Based on data obtained with the European
Southern Observatory Very Large Telescope, Paranal, Chile, program
070.A-9007(A), and on data obtained at the Canada-France-Hawaii
Telescope, operated by the CNRS of France, CNRC in Canada and the
University of Hawaii.}} 

\section{Introduction}

Early-type galaxies are the preferred target for studies on how and
when galaxies were formed, because of the simpler task of modeling an
old stellar population undergoing passive evolution compared to
modeling a young population continuously modified by an irregular star
formation history.  Over the last few years, mostly in response to the
accumulating evidence in favor of a predominantly old stellar
population within early type galaxies \citep[see for a complete
review][]{renzini_review}, the discussion has focused onto two main
areas: the history of stellar formation and how and when stars
assembled into galaxies. Still, a fundamental pre-requisite for such
studies is to isolate from cosmological surveys a sample of early-type
galaxies representative of the true galaxy population at all epochs
probed.

Commonly used galaxy classification schemes are based purely on the
morphological appearance of galaxies. It is well known however that
galaxies follow a number of scaling relations involving their stellar
populations and their morphological, structural and photometric
parameters.  Compared to their late-type counterparts, early-type
galaxies have been found to be redder 
\citep[color-morphological type relation,][]{color_morph}, more luminous 
\citep[color-magnitude relation,][]{color_mag1,color_mag2}, to be 
located in denser environments 
\citep[morphology-density relation,][]{dens_morph}, and
to have a star formation history that takes place over shorter
time-scales \citep{sandage,gavazzi_cri}.

Cosmological survey studies, faced with the difficulty of obtaining a
morphological classification for all the galaxies in their sample,
have often tried to take advantage of those relations to define an
alternate classification scheme. Galaxy color, which is by far the
easiest parameter to measure for a full survey sample, has been most
commonly used as a substitute for morphological information. Lately
this practice has become even more commonplace, since the galaxy
rest-frame color distributions have been found to be clearly
bimodal. This bimodality was well known to exist within clusters of
galaxies, where the early- and late-type galaxies follow two rather
distinct color-magnitude relations, as discussed by \citet{color_mag1}
and \citet{local_cmr} for early-type galaxies and \citet{color_mag2}
and \citet{gavazzi_mass} for late-type ones. That the same bimodality
was present in a general field sample was first noticed in the local
universe by \citet{bimodal_1} using the SDSS galaxies sample, and
afterwards discussed by many other authors \citep[see, as an
example,][]{bimodal_8,bimodal_9}.  \citet{bimodal_bell} (hereafter
B04) and \citet{bimodal_6} detected the rest-frame color bimodality
also at higher redshifts up to z$\sim$1, respectively in the COMBO-17
and DEEP2 data.

Color bimodality is interpreted as just one specific signature
introduced by the general processes controlling galaxy formation and
evolution \citep{bimo_models,bimo_models_2}.  Other similar signatures
are observed, since bimodality has been found to be a characteristic
in the distribution of many other observables like H$_\alpha$ emission
\citep{bimodal_ha}, 4000 $\AA$ break \citep{bimodal_4000}, star
formation history \citep{bimodal_sfh}, or clustering
\citep{bimodal_clustering,vvds_clustering}.

Because of the age of their stellar population, early-type galaxies
are expected to have very red optical colors over a rather large
redshift range.  This expectation has been confirmed in clusters of
galaxies, where the red-sequence for morphologically selected
early-type galaxies has been observed up to z$\sim$1.2
\citep{kodama_models,mono_test_4}. As a result the red color has often
been used as a tracer to isolate samples of early-type galaxies. Most
recently the widespread acceptance of color bimodality has resulted in
the use of red-sequence galaxies (those that populate the red peak of
the color bimodal distribution) to study the evolution of early-type
galaxies, implicitly assuming this red population to be entirely
composed of old, passively evolving objects. 

However, while most of the early-type galaxies are indeed red, they
are possibly not the only red objects included in a survey
sample. Attempts at quantifying the contamination of non early-type,
passive objects within samples of red galaxies include both studies
focused on Extremely Red Objects (EROs), and more general surveys
which target the whole bulk of the galaxy population. Among EROs,
\citet{cimatti_eros}, using spectroscopic data for 45 objects, find a
roughly equal proportion of early-type, passive galaxies and of dusty
starburst objects. Among less extreme objects, in the local Universe
\citet{bimodal_1} reported a significant fraction (20\%) of galaxies
morphologically later than Sa in the red galaxy population. This
result is confirmed using the data for galaxies in the Virgo Cluster
and in the Coma Supercluster provided by the GOLDMine database
\citep{goldmine}. At intermediate redshift (up to $z\sim1$) various
studies are confirming these results \citep[see for
example][]{redpop_hdf,redpop_gems}, while the nature of red galaxies
at higher redshift is less clear. However, there are indications that
the fraction of morphologically late-type galaxies with red colors
could be even higher at $z\gtrsim~1$ \citep{bimodal_1}, in agreement
with the findings based on EROs studies.

In this work we use VIMOS-VLT Deep Survey photometric and
spectroscopic data \citep[VVDS, see][]{vvds_main} to compare the
galaxy population that can be isolated by using either a red color
selection criterion or a spectro-photometric classification.  Although
neither of these selection criteria can be considered entirely
equivalent to a morphological classification in isolating a sample of
early-type galaxies, our comparison does provide an estimate for the
uncertainties involved in selecting those objects within a
cosmological survey sample.

The paper is organized as follows. In section \ref{sec:sample} we 
describe the VVDS galaxy sample, the data we use in this work and we
discuss the effect that the VVDS selection function could have on
our work. In section \ref{sec:bimo} we analyse the 
rest-frame color distributions and we demonstrate that the color 
bimodality is present up to at least z=1.5. Then we study the color 
bimodality as a function of redshift and of galaxy luminosity.  
In section \ref{sec:cmr_red} we focus on the red-sequence
galaxies studying their color-magnitude relation. In section
\ref{sec:d4000_OII}, we analyse in detail the red-sequence objects
showing how this population is contaminated by a non negligible
fraction of starforming objects. We use the strong bimodality
observed in the EW([OII])-D$_n$(4000) distribution to isolate a sample
of early-type galaxies less contaminated by star-forming galaxies than
a simple color-magnitude selection.  
Finally in section \ref{sec:cmr_early} we discuss the color evolution 
of this sample.

All magnitudes are given in the Johnson-Kron-Cousin system; the
adopted cosmology is the standard $\Omega_m=0.3$,
$\Omega_{\Lambda}=0.7$ and $H_0=70~km~s^{-1}~Mpc^{-1}$.

\section{Sample description}
\label{sec:sample}

\subsection{The data}
\label{sec:data}

\subsubsection{Observations}

Our sample is selected from the first epoch VVDS-Deep spectroscopic
sample within the VVDS-0226-04field (hereafter F02) \citep[see][]{vvds_main}. 
This is derived from a purely magnitude limited sample (hereafter 
``photometric sample''), including all objects in the magnitude range 
17.5 $\le I_{AB} \le$ 24.0 from a complete, deep photometric survey 
\citep{vvds_imaging}.

The whole 1.2 deg$^2$ field has been imaged in $B, V, R$ and $I$ with
the wide-field 12K mosaic camera at the Canada-France-Hawaii Telescope
(CFHT), reaching the limiting magnitudes of $B_{AB} \sim 26.5, V_{AB}
\sim 26.2, R_{AB} \sim 25.9, I_{AB} \sim 25.0$. Data are complete and
free from surface brightness selection effects down to $I_{AB} \le
24.0$ \citep[see for details][]{vvds_imaging_f02}.  U-band data, taken
with the Wide-Field Imager at the ESO/MPE 2.2m telescope, are
available for a large fraction of the field, to the limiting magnitude
of $U_{AB} \sim 25.4$ \citep{vvds_imaging_u}. 
In all bands, apparent magnitudes have been measured using Kron-like 
elliptical apertures (the same in all bands), with a minimum Kron 
radius of 1.2 arcsec, and corrected for the Galactic extinction using 
the Schlegel dust maps \citep{dust_map}. The median extinction correction
in the $I$ band is $\sim 0.05$ magnitudes.

Spectroscopic observations for about 23\% of the objects included in
the photometric magnitude limited sample have being carried out with
the VIsible Multi Object Spectrograph \citep[VIMOS, see][]{vimos_comm}, 
on the UT3 unit telescope of the ESO Very Large Telescope. 
The spectroscopic sample is close to be a perfectly random
subset of the parent photometric sample, with only a small bias
against large angular-size objects which can be easily corrected for
\citep[see][]{vvds_lumfunc,vmmps}.

Observations have been reduced with the VIMOS Interactive Pipeline 
and Graphical Interface software package
\citep[VIPGI, see][]{vipgi,ifu}. A redshift measurement has been
derived for 9036 galaxies % using the KBRED software package \citep{kbred}, 
with a success rate which is mildly dependent on both
the galaxy apparent magnitude and its redshift \citep[see the
discussion on the Spectroscopic Sampling Rate and Figures 2 and 3
in][]{vvds_lumfunc}. The median redshift for the whole spectroscopic
sample is $z=0.76$ \citep[see Figure 25 in][]{vvds_main}.

The sample we use for this work is composed of all the galaxies with
secure identification and redshift measurement \citep[z quality flags
2,3,4,9 in][]{vvds_main} in the F02 spectroscopic sample, with the
restriction of having $z\leq 2.0$, in order to ensure that the
observed optical and near-infrared magnitudes provide a reasonably
close bracketing for the rest-frame optical magnitudes we use in our
analysis. A total of 6291 galaxies are included in this sample with a
measured B, V, R and I magnitude; hereafter we will refer to this sample
as ``complete sample''.

\subsubsection{Rest-frame absolute magnitudes and colors}

Rest-frame colors for all galaxies in the ``complete sample'' are 
computed from the absolute magnitude estimates derived as described 
in Appendix A of \citet{vvds_lumfunc}. 
For a given rest-frame photometric band and a
given galaxy redshift, the absolute magnitude is obtained from the
apparent magnitude measurement within the observed photometric band
that most closely matches the rest-frame one. A k-correction is then
applied to take into account the residual difference between the two
photometric bands. The amplitude of the k-correction is derived by
selecting the best-fitting galaxy spectra template to the galaxy
B, V, R, I magnitude measurements.
In this work we focus on the rest-frame U-V color, mainly because it
allows us to sample the amplitude of the 4000 \AA~ break in the
Spectral Energy Distribution, which is a good indicator of the galaxy
stellar population age \citep{bruzual83,ste_mass_1}. U-V is also a 
color often used by previous works on the properties of early-type 
galaxies, both locally \citep[for example][]{color_mag1,local_cmr} 
and at redshifts up to 1 (see B04).
Another advantage provided by this choice of rest-frame photometric 
bands is that they are bracketed by the observed bands for most of 
the redshift range covered by our sample. As discussed in 
\citet{vvds_lumfunc}, typical k-correction uncertainties are of the 
order of 0.04 mag for the U-band, and of 0.11 mag for the V-band.  
When these uncertainties are added to the apparent magnitude 
measurement one (approximately 0.1 mag for the faintest object), 
they contribute to a total color measurement uncertainty of up 
to 0.2 mag.

\subsubsection{Spectral indexes}

For all objects in the ``complete sample'' we have obtained measurements of 
the equivalent width and line flux for all emission lines and absorption
features detected in their spectra. 
All these measurements have been carried out using the PLATEFIT software 
package (Lamareille et al. 2007, in preparation), which implements the 
spectral features measurement techniques described by \citet{mass_metallicity}. 
In this work we use exclusively measurements of the [OII]$\lambda$3727
doublet equivalent width and of the 4000 \AA\ break (D4000), using the 
narrow break definition proposed by \citet{balogh_spectral_indexes}.
Because of the VVDS observational set-up, these measurements are 
possible only for objects in the redshift range $0.45<z<1.2$, i.e. 4433 
objects \citep[see][]{vvds_main}: hereafter we will refer to this sample
as ``spectroscopic sub-sample''.

\subsection{The VVDS selection function} 

To discuss in a meaningful way galaxy colors, their evolution, and their
relation to other physical properties of galaxies, we must consider possible
biases against the inclusion of galaxies of a given (extreme) color in 
our ``complete sample''. Any such bias could be introduced either by the
definition of the magnitude limited ``photometric sample'' or by some bias in
the efficiency of the redshift measurements used to define our 
``complete sample''.

\subsubsection{Limiting magnitude} 
\label{sec:magbias}

Within the VVDS, as in all magnitude limited surveys, the range of
luminosities covered by the sample changes with redshift, with a global trend
towards higher luminosities with increasing z. The fixed magnitude limit
however is also introducing some color bias at the faint luminosity end of the
sample. As the apparent magnitude for a galaxy in a given photometric band
depends on its luminosity, redshift and spectral energy distribution (and
therefore on its color), it is a expected that, for a given redshift and
luminosity, galaxies with rest-frame color above a certain value should be
systematically excluded from the sample, simply because their apparent I-band
magnitude becomes too faint, while bluer galaxies with the same luminosity are
still brighter than the limiting magnitude because of their flatter spectrum
\citep[see][]{lumfunc_bias}. To quantify how this selection effect might bias
our results we have divided our ''complete sample'' in four redshift and absolute 
magnitude bins and derived in each bin a rest-frame color vs. apparent 
I-band magnitude relation using synthetic spectral energy distributions (SED). 

The SEDs were obtained using publicly available stellar population 
synthesis models from Bruzual \& Charlot \citep{bc_last}, which provide 
the time evolution of a galaxy stellar spectrum as a function of galaxy age, 
star formation history and stellar initial mass function. 
For this work we have always used a Salpeter initial mass function 
\citep{salpeter}. For the star formation history (SFH) we follow 
\citet{gavazzi_cri} in using a slightly more realistic set of SFHs with 
respect to the commonly used exponentially decreasing one. To cover as 
wide a range of galaxy properties as possible, we generated a wide array
of synthetic SEDs (ages from 0 to 15 Gyr and star formation time-scales 
from 0.1 to 25 Gyr). 

Because of the non negligible ranges of redshift and luminosity spanned by
each of our redshift and absolute magnitude bins we simulate a uniform 
distribution of galaxies inside the bin, assign to each of them one SED
and use them to derive the joint distribution of apparent I-band magnitude 
and rest-frame U-V color.

By imposing to this distribution the same apparent magnitude limit used to
select the VVDS sample we can measure the likelihood for a galaxy with a
given color to be included in our ``complete sample'' which we term 
``color completeness''. 
Whenever this completeness is too low (say $\lesssim$ 50\%) we consider 
the corresponding color strongly biased against in the bin.
 
The assumption of a uniform luminosity and redshift distribution for the 
objects within each bin is not completely realistic. However within the two
high-redshift bins, where the bias can be non-negligible, the 
larger fraction of fainter objects (more heavily biased against, because they 
are less likely to be observable above the sample apparent magnitude limit) 
is compensated quite effectively by a larger fraction of objects in the 
lower half of the redshift bin (less biased against, because their smaller 
distance makes them more likely to be observable). 

In section \ref{sec:bimo} we apply this estimate of the color bias introduced 
by the VVDS selection function to show that within the adopted redshift 
bin limits ($0.2, 0.6, 0.8, 1.2, 2.0$) and $V$-band absolute magnitude
bin limits (-22, -21, -20, -19) the bias is mostly neglibgible, and does not 
significantly affects the observed color distributions.

\subsubsection{Redshift measurement efficiency}
\label{sec:redbias}

Another possible source of bias for our data could be the systematic
loss of early-type galaxies within our ``complete sample''.
This loss could be due either to the arguably lower efficiency
in measuring spectroscopic redshifts for early-type with respect 
to late-type galaxies or to a possible bias against observations 
of early-type objects in multi-object spectroscopic surveys like
the VVDS.
This bias originates from the stronger clustering of these objects
with respect to late-type ones, which cannot be matched because of
MOS mask design limitations. 

To quantify this possible bias we have used the ``photometric type'' classification
scheme proposed in \citet{lum_type_vvds}; the available optical photometry
data for the whole photometric magnitude limited sample have been fitted
with local spectral templates taken from \citet{cww}, supplemented with 
two starburst templates, to derive a photometric type for each galaxy in 
our sample. Here we are considering the E/S0 and the early spiral photometric 
type objects together as a ``photometric early-type population'', and the 
late-type spiral, irregular and starburst photometric type objects 
together as a ``photometric late-type population''.

We have measured the percentage of photometric early-type objects in the whole 
``photometric sample'' and in our spectroscopic ``complete sample''. These values 
are listed in table~\ref{tab_rme} for the whole redshift range and for two
subsamples at low and high redshift. In this case we consistently used for 
both samples photometric redshifts, determined as described in \citet{photoz_cfhtls}.
By comparing the fraction of early-type galaxies in the two samples
we estimate that the missing early-type objects in our spectroscopic
''complete sample'' are very few ($\sim 4\%$ of the total number of objects).

\begin{table}[h!]
\begin{center}
\caption{The percentage of photometric early-type objects in the whole 
``photometric sample'' (``Photo'' columns) and in our ``complete sample''
(``Spec'' columns). Values are listed for the whole redshift range and 
for two redshift bins.  
}

\label{tab_rme} \[
\begin{tabular}{c|cc|cc|cc}
\hline
\hline
\noalign{\smallskip}
Type & \multicolumn{2}{c}{All} & \multicolumn{2}{c}{$0.2<z<0.8$} & \multicolumn{2}{c}{$0.8<z<2.0$} \\
     & Photo & Spec &  Photo & Spec &  Photo & Spec    \\
\noalign{\smallskip}
\hline
\noalign{\smallskip}

Early &  29.1 &  26.6 &  30.3 &  28.5 & 29.0 &  24.5 \\
Late  &  70.9 &  73.4 &  69.7 &  71.5 & 71.0 &  75.5 \\

\noalign{\smallskip}
\hline
\end{tabular} \]
\end{center}
\end{table}

\section{Rest-frame color bimodality} 
\label{sec:bimo}

\begin{figure*}
\resizebox{\hsize}{!}{\includegraphics[clip=true]{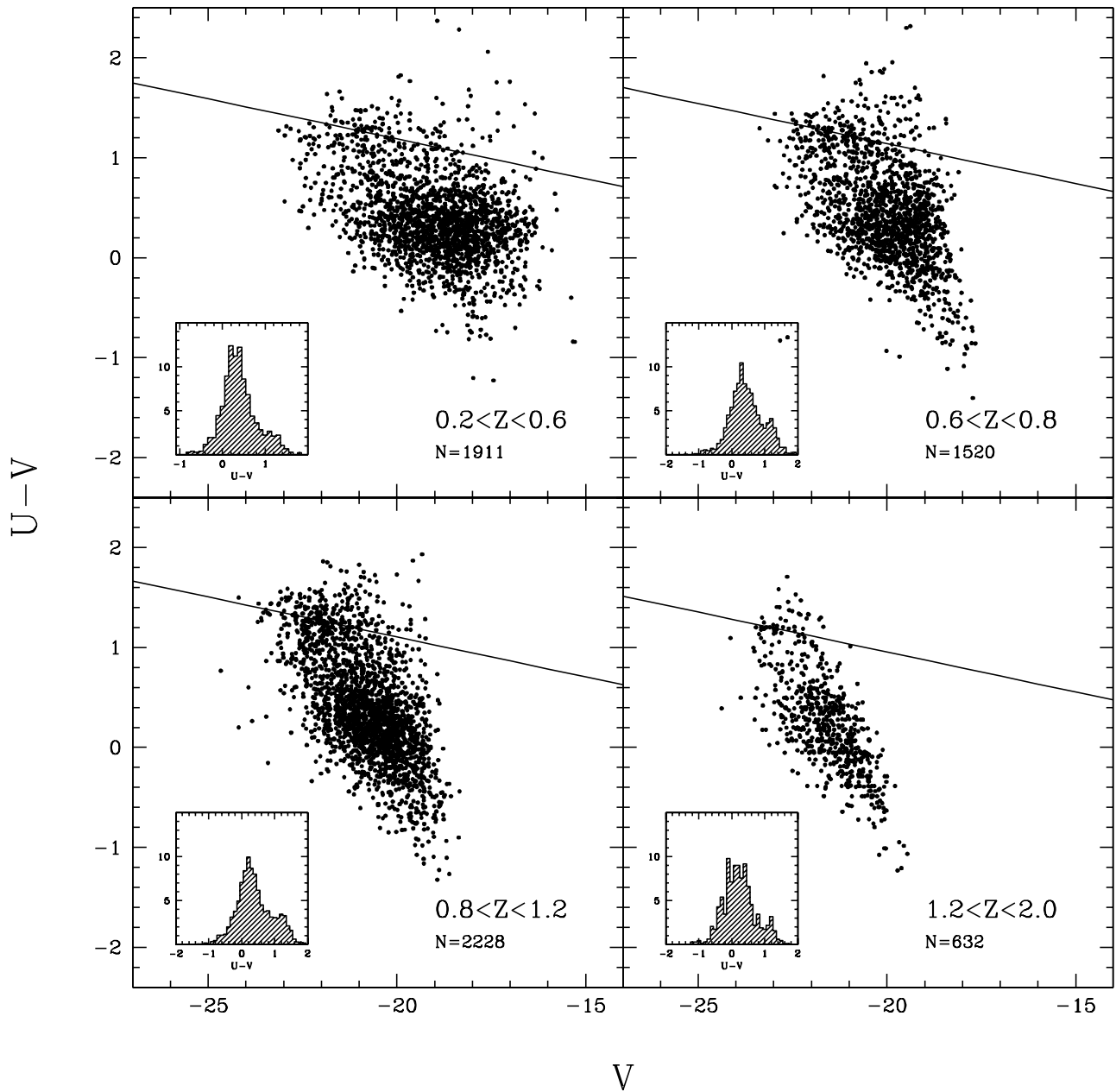}}
\caption{\footnotesize
The rest-frame U-V color against the absolute V magnitude for our spectroscopic
``complete sample''. The four panels show galaxies in 
different redshift intervals, as indicated in the panels themselves. 
The solid line in each panel shows the best fitting color-magnitude relation
for the red sequence objects within the given redshift interval 
(see section \ref{sec:cmr_red}).
The inset in each panel shows the color distribution for all the objects in
that redshift interval, i.e. the projection of the color-magnitude relation on  
the color axis; notice that we plot here percentages and not absolute numbers 
of objects. 
}
\label{colmag}
\end{figure*}

\begin{figure*}[t!]
\resizebox{\hsize}{!}{\includegraphics[clip=true]{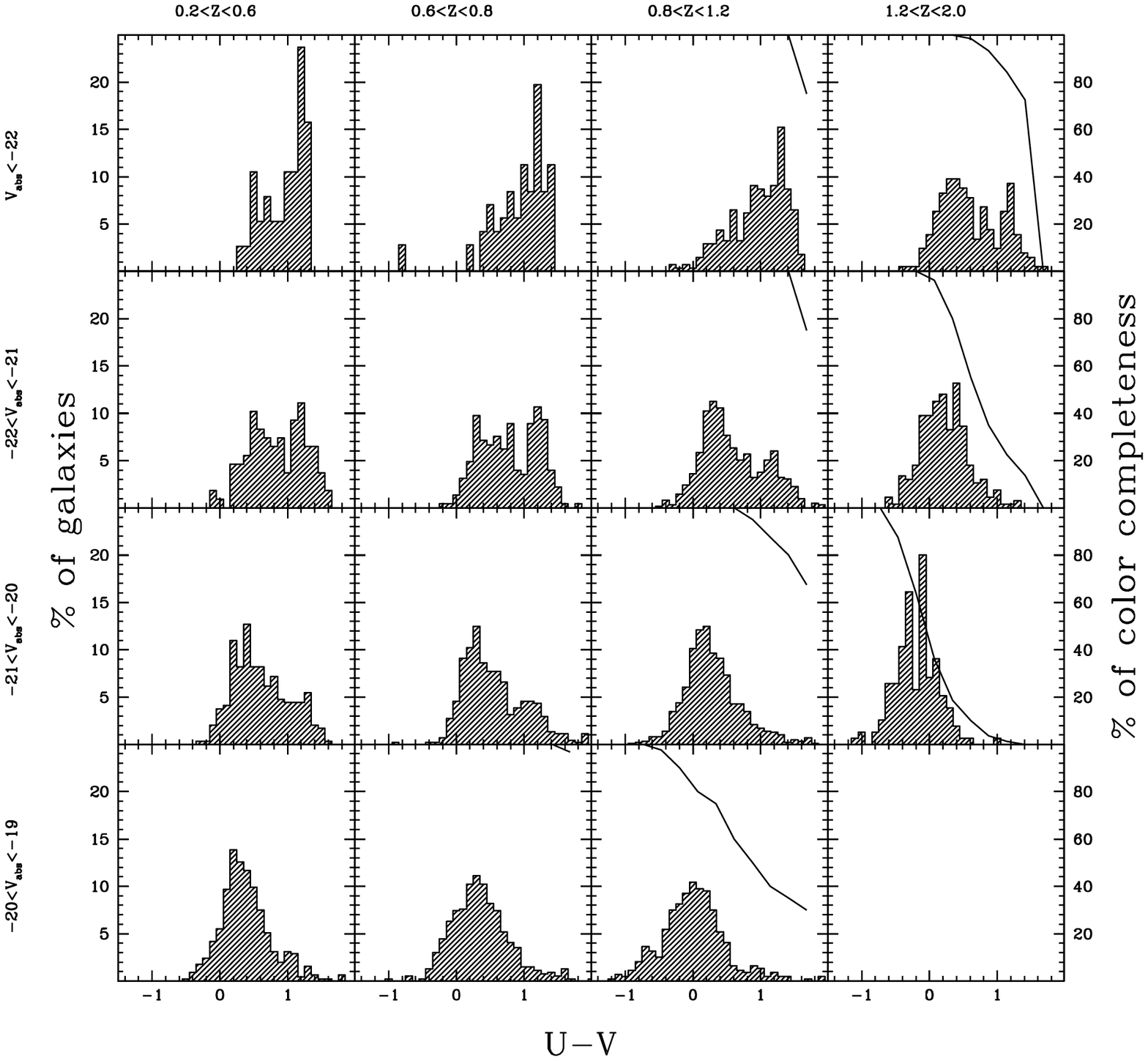}}
\caption{\footnotesize
The U-V rest-frame color distributions for our spectroscopic ``complete
sample'' divided into four V mag absolute magnitude and four redshift bins;
luminosity increases from the bottom to the top, redshift increases from left
to right. The histograms scale is shown on the left side of the plot.  The
solid line plotted in some panels highlights the color regions which are
biased against by the VVDS selection criteria (see paragraph \ref{sec:magbias}
for details); the scale for these lines, i.e. the fractional bias against a
given color value in the sample, is shown on the right side of the plot.}
\label{mosaic}
\end{figure*}

Color bimodality has been observed for galaxy samples from many different
surveys, and the VVDS is no exception in this respect.
The distribution of the rest-frame U-V color against the absolute V magnitude
for our ``complete sample'' is shown in figure~\ref{colmag},
with the total sample divided into four different redshift intervals. The
small inset in each panel shows the corresponding color distribution, i.e. the
projection of the color-magnitude relation on the U-V axis.
A bimodal color distribution is observed over all four redshift intervals.
We remark that within our data the bimodality is visible irrespective of the
detailed choice of filters that one could use to define the rest-frame color:
we observe a rather similar bimodal distribution using any combination of the
U, B, V, R, and I filters.  The observed bimodal color distribution is in
agreement with previous SDSS \citep{bimodal_1}, DEEP \citep{bimodal_6} and
COMBO-17 (B04) findings.  Moreover the depth of our sample is allowing us to
extend the detection of such a bimodality up to at least $z=1.5$ (the mean
redshift for the galaxies in the highest redshift bin). This is in good
agreement with the galaxy formation model discussed recently by
\citet{bimo_models}, who predict a color bimodality to be clearly present 
starting from $z\approx1.5$ (see their figure~4), as a result of the interplay
between the merging histories of forming galaxies and the feedback/star formation
process. 
 
The insight that the bimodal rest-frame color distributions provide 
into galaxy formation and evolution processes is however limited, mainly
because this bimodality is observed only when objects of all luminosities 
are considered at once. 
The fact that both early- and late-type galaxies
follow some form of color-magnitude relation, although the two
relations are quite different from each other, has certainly an impact 
on how the red and blue components in the color distribution presented 
above are populated. 
Equally important in this respect is the role of the environment, driven by
the density-morphology relation, and the work  of \citet{bimodal_9} on
the SDSS sample illustrates the gradual change in the proportion of
blue and red galaxies as a bivariate function of galaxy luminosity and
local density. \citet{vvds_density} have obtained similar results on
the VVDS data. 

The influence that luminosity has on the proportion of red and blue galaxies
is particularly important in our study, because of the large redshift interval
covered by our sample, and the variation in the range of luminosities that are
spanned at different redshifts by a magnitude-limited sample like the VVDS
one. In figure~\ref{mosaic} therefore we further subdivided our ``complete
sample'', by splitting it within each redshift bin showed in
figure~\ref{colmag} into four V-band absolute magnitude bins.  In
agreement with the results of \citet{bimodal_9}, the global bimodal color
distribution is replaced mostly by unimodal distributions, whose properties 
depend both on redshift and on galaxy luminosity.  It is evident that a global
color-magnitude relation is present over the whole $0<z<2$ redshift interval,
with galaxies becoming redder as their luminosity increases.  At the same time
we also notice how, within a fixed luminosity bin, galaxies become bluer with
increasing redshift, and this effect is present at all luminosities covered by
our sample.

As discussed in section \ref{sec:magbias}, this trend towards bluer colors
could be, in principle, partly due to a sample selection bias, introduced by
the fixed magnitude limit used to define the VVDS sample. To demonstrate that
in practice this is not the case, within each redshift-absolute magnitude panel
we indicate with a thick solid line our estimate of the 
color completeness. Whenever this completeness is low ($\lesssim$ 50\%, see 
discussion in paragraph \ref{sec:magbias}) the corresponding color is
clearly biased against in our sample, and therefore the color 
distribution plotted for the given redshift and luminosity bin might
not be representative of the whole galaxy population within the bin
boundaries. From the figure we can clearly see that the color completeness is
good, with the partial exception of the $1.2<z<2.0$ and $-21<V_{abs}<-20$ bin, 
demonstrating  how the visible trend towards bluer color at high redshift for 
all luminosities is not artificially created by the VVDS sample definition.

These results are in agreement with the findings of previous redshift surveys 
like the CFRS \citep{cfrs_iii} and the Hawaii Deep Fields Survey \citep{hdfs}, 
but it is the first time that these findings are extended significantly
over the $z\approx1$ limit.

\section{The color-magnitude relation of the red galaxies}
\label{sec:cmr_red}

\begin{figure}[t!]
\resizebox{\hsize}{!}{\includegraphics[clip=true]{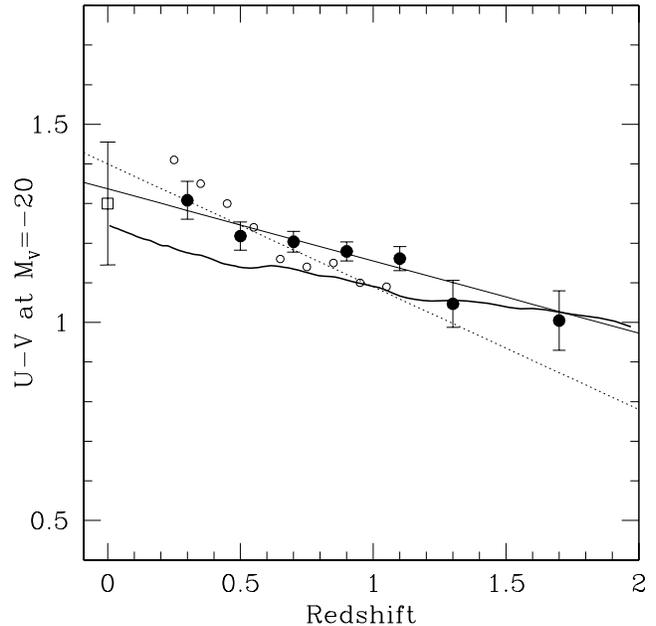}}
\caption{\footnotesize
Color evolution of the red sequence as measured by the value of the
intercept of the CMR at $M_V=-20$. Small open circles and the dotted
line are from B04; large filled circles and the solid line represent
our VVDS data and the best linear fit to them. The open square point
at $z=0$ is the $U-V$ CMR zero point at $z=0$ computed by B04 from
SDSS data. The thick solid curve shows the color evolution of the
template which, in our grid, better approximates a Single Stellar
Population (a single burst 0.1 Gy long at z=5, followed by pure passive 
evolution), and it is not a fit to the observed evolution of the color
of the red sequence. 
Error bars on our points account only for the statistical uncertainties
in the CMR zero-point determination}
\label{red_evol_1}
\end{figure}

\begin{table}[h!]
\caption{The red-sequence CMR intercept and dispersion}
\begin{center}
\label{tab_cmr} \[
\begin{array}{cccc}
\hline
\hline
\noalign{\smallskip}
Redshift & N galaxies & (U-V)_{M_V=-20} & \sigma \\
\noalign{\smallskip}
\hline
\noalign{\smallskip}
0.2-0.4   &       82 &    1.31 &  0.31 \\ 
0.4-0.6   &      123 &    1.22 &  0.25 \\ 
0.6-0.8   &      259 &    1.20 &  0.29 \\ 
0.8-1.0   &      229 &    1.18 &  0.21 \\ 
1.0-1.2   &      133 &    1.16 &  0.18 \\ 
1.2-1.4   &       44 &    1.05 &  0.25 \\
1.4-2.0   &       17 &    1.00 &  0.08 \\
\noalign{\smallskip}
\hline
\end{array} \]
\end{center}
\end{table}

While in the previous section we analyzed the color distribution for the whole
spectroscopic ``complete sample'', here we focus our attention on the galaxies
in the read peak of the bimodal distribution, under the temporary assumption
that these red objects can be identified with the early-type galaxy population
within our sample.  The properties of the color-magnitude relation (CMR) of
early-type galaxies (scatter, slope, and evolution with redshift) have often
been used to constrain formation and evolution scenarios for this galaxy
population \citep[see for example][]{local_cmr, kodama_models,bernardi03}.
These studies were mostly based on cluster samples of morphologically defined
early-type galaxies, and it was not entirely obvious how a similar kind of
analysis could be performed on a sample of color-selected objects in a large
high redshift survey lacking the necessary high resolution imaging to perform 
the appropriate morphological classification.
B04 have demonstrated that with a large sample and good photometric redshift
estimates like those provided by the COMBO-17 survey, such kind of analysis is
indeed possible. Here we follow their procedure to study the redshift
evolution of the red CMR in the VVDS sample.

Following B04, we have used our data to estimate the zero-point of the CMR,
keeping its slope fixed to the value which has been determined for local
galaxy clusters \citep[-0.08, see][]{local_cmr}. Within each redshift bin 
we first corrected the individual color measurements to eliminate the fixed CMR slope,
and then we used the bi-weight estimator \citep{biweight} to compute the mean
color for all red galaxies, defined (as in B04) as objects with rest-frame color U-V$>$1.0. 
The solid line plotted in each panel of figure~\ref{colmag} shows the
corresponding fit for the red sequence in that redshift range.

The change in the value of the fitted CMR intercept at $M_V=-20$ as a function
of redshift is shown in Figure~\ref{red_evol_1}, and the plotted values are
also summarized in Table~\ref{tab_cmr}. The evolution of the red population is quite 
visible; from $z\sim0.3$ to $z\sim1.7$ red galaxies
become bluer on average by $\sim0.3$ mag. The solid line is a linear fit to the
evolution: $U-V=1.355-0.206 \times z$.  The results of B04 on the COMBO-17
data are plotted for comparison; also the $U-V$ CMR zero point at $z=0$
determined by B04 from SDSS data is shown. Our results qualitatively confirm
the evolution claimed by B04 and extend their measurement up to $z\sim2$.
However we find that the amount of this evolution, quantified by the slope of
the relation, is somewhat milder than that reported by B04, and this result is
bringing the whole evolutionary trend in better agreement with the $z=0$
SDSS-based point. The thick line plotted in figure~\ref{red_evol_1} shows,
purely as a reference, the color evolution of the synthetic SED within the
models grid discussed in section~\ref{sec:magbias} which better approximates a 
Single Stellar Population (a single burst 0.1 Gy long at z=5, followed by pure 
passive evolution). As already noticed by B04, there is
a general good agreement between the observed evolution of the CMR and the
expectations for a purely passively evolving population. Our results show that
this agreement extends at least up to $z\sim2$.

\begin{figure*}
\begin{center}
\resizebox{\hsize}{!}{\includegraphics[clip=true]{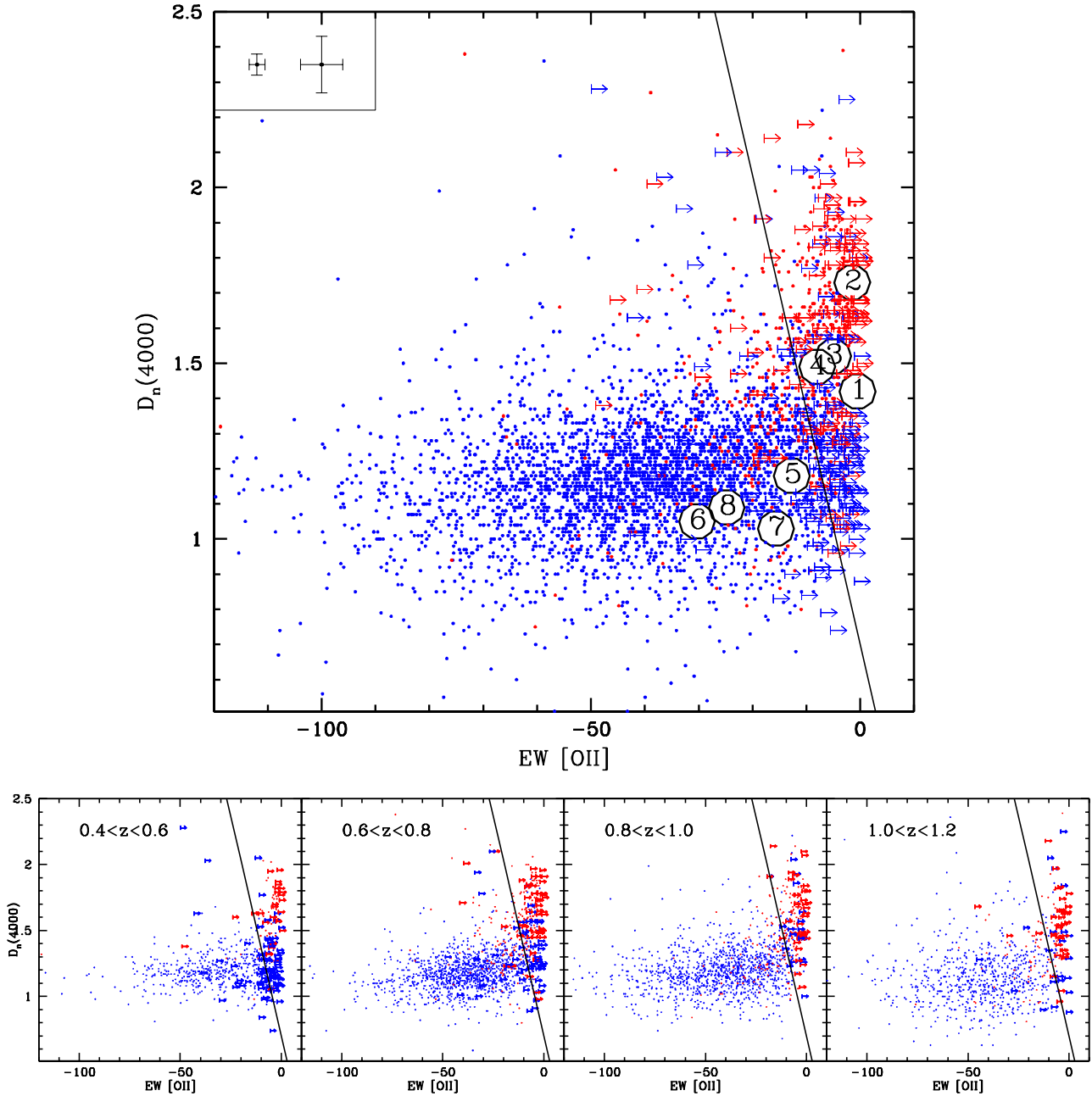}}
\caption{\footnotesize
The D$_n$(4000) distribution as a function of the rest frame
EW[OII] for the whole ``spectroscopic sub-sample'' (top
panel) and for the same sample subdivided into the four 
indicated redshift bins (bottom panels). 
Blue symbols are objects with $U-V<1.0$, while red symbols are 
objects with $U-V>1.0$. Only 2$\sigma$ or higher confidence D$_n$(4000)
measurements are plotted. For EW[OII] measurements with a
confidence level lower than 2$\sigma$ upper limits are plotted The
error bars in the upper left corner of the plot show the typical
measurement uncertainties for the higher S/N spectra (left) and the lower 
S/N spectra (right) sub-samples.  The solid line divides the sample in the 
passive population (ET) and star-forming population (LT). For comparison the
large numbered circles are the D$_n$(4000) / EW[OII] values for the
Virgo cluster templates from \citet{gavazzi_cri} as listed in Table
\ref{tab_virgo} }
\label{d4oii}
\end{center}
\end{figure*}

\begin{table}[]
\caption{The D$_n$(4000) and rest-frame EW([OII]) values for the Virgo templates plotted in 
figure~\ref{d4oii}}
\begin{center}
\label{tab_virgo} \[
\begin{array}{cccc}
\hline
\hline
\noalign{\smallskip}
{\rm Label} & {\rm Type} & {\rm D_n(4000)} & {\rm EW([OII]) (\AA)} \\
\noalign{\smallskip}
\hline
\noalign{\smallskip}
1 &  {\rm dE}   &  1.42  & > -0.5   \\ 
2 &  {\rm E/S0} &  1.73  & > -0.5   \\ 
3 &  {\rm Sa}   &  1.52  & > -5.0   \\ 
4 &  {\rm Sb}   &  1.49  & > -8.0   \\ 
5 &  {\rm Sc}   &  1.18  & -12.7   \\ 
6 &  {\rm Sd}   &  1.05  & -30.3   \\ 
7 &  {\rm Irr}  &  1.03  & -15.7   \\
8 &  {\rm BCD}  &  1.09  & -24.8   \\
\noalign{\smallskip}
\hline
\end{array} \]
\end{center}
\end{table}

\section{The nature of the red population: separating star-forming 
from passively evolving galaxies}
\label{sec:d4000_OII}

\subsection{Spectral properties}

Early-type galaxies are dominated by an old stellar population undergoing an
almost purely passive evolution, and this last property is what makes them an
attractive target for galaxy evolution studies. Therefore, in selecting
early-type galaxies from a global survey sample we should consider how to
efficiently select passively evolving objects, and not just red galaxies or
galaxies that are morphologically classified as elliptical of S0.

As a first step towards this goal, we start by quantifying how the
contamination from late-type, star forming galaxies is affecting the
properties of the red color-selected population, and how this effect is
changing with redshift. We use the VVDS spectroscopic information to separate
our ``spectroscopic sub-sample'' into two classes of old, most likely 
passively evolving and of young, star--forming objects. We remind that
this sample is restricted to the redshift range $0.45<z<1.2$ (see
paragraph \ref{sec:data}); all the analysis done in the following sections 
is therefore limited to this redshift range.
A more detailed analysis of the spectral properties of this sample will
be presented in Vergani et al. 2007 (in preparation).

In figure~\ref{d4oii} (top panel) the relation between the equivalent 
width of the [OII]$\lambda$3727 line (EW[OII]) and the 4000 
\AA~break (D4000) is shown for the whole ``spectroscopic sub-sample''.
Blue symbols are objects with $U-V<1.0$, while red symbols are objects 
with $U-V>1.0$.
It is clearly visible from the plot how objects follow a bimodal L-shaped
distribution, with most of the star-forming objects that have a large
EW[OII] having a negligible D4000, and vice-versa most of the objects
that have a strong D4000 showing no significant [OII] emission in
their spectra. 
The bottom panels of figure~\ref{d4oii}, obtained dividing the  
``spectroscopic sub-sample'' in four redshit bins, show that
this bimodality is essentially independent from redshift.
The larger dispersion in the D4000 measurements
of star-forming objects in the higher redshift bin is mostly due to 
the higher uncertainties in measuring this feature for distant and 
faint objects.

These distributions suggest a natural sub-division of the galaxy
population into two classes. We define \emph{spectral early-type}
objects (ET) galaxies in the vertical arm of the two parameters
distribution, i.e. those that do not show any detectable sign of star
formation activity, and which can be expected to undergo further
evolution only via passive evolution of their stellar
population. Conversely, we define \emph{spectral late-type} objects
(LT) galaxies in the horizontal arm, still undergoing a vigorous star
formation activity.  More elaborated spectral classification schemes
have been proposed in the past, like the four-fold subdivision
discussed by \citet{k20_spectro}, but we prefer to use here a simpler
two-fold subdivision, that mirrors the one obtained using the
rest-frame colors.

\begin{figure*}
\begin{center}
\resizebox{8.5truecm}{!}{\includegraphics[clip=true]{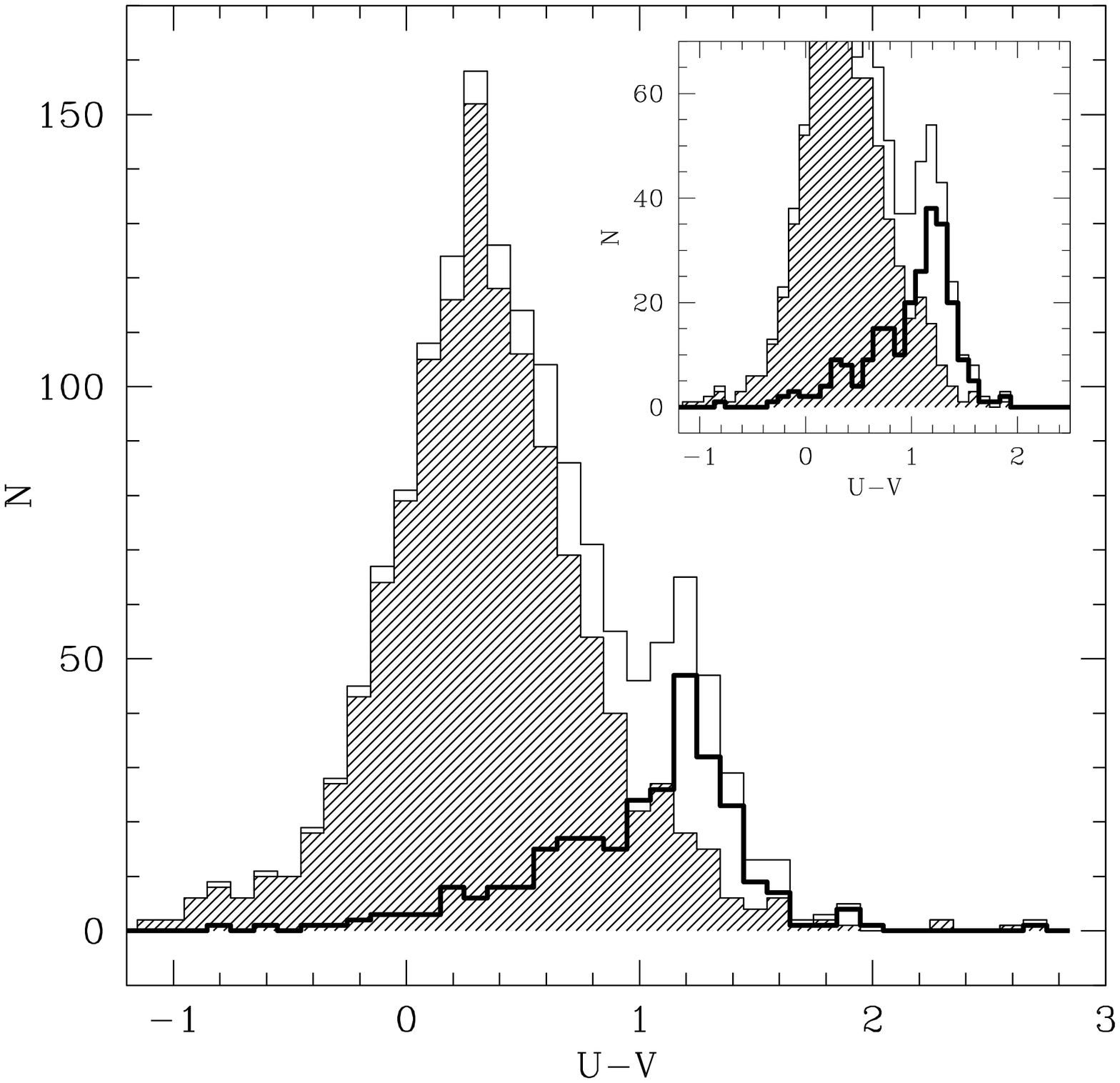}}
\resizebox{8.5truecm}{!}{\includegraphics[clip=true]{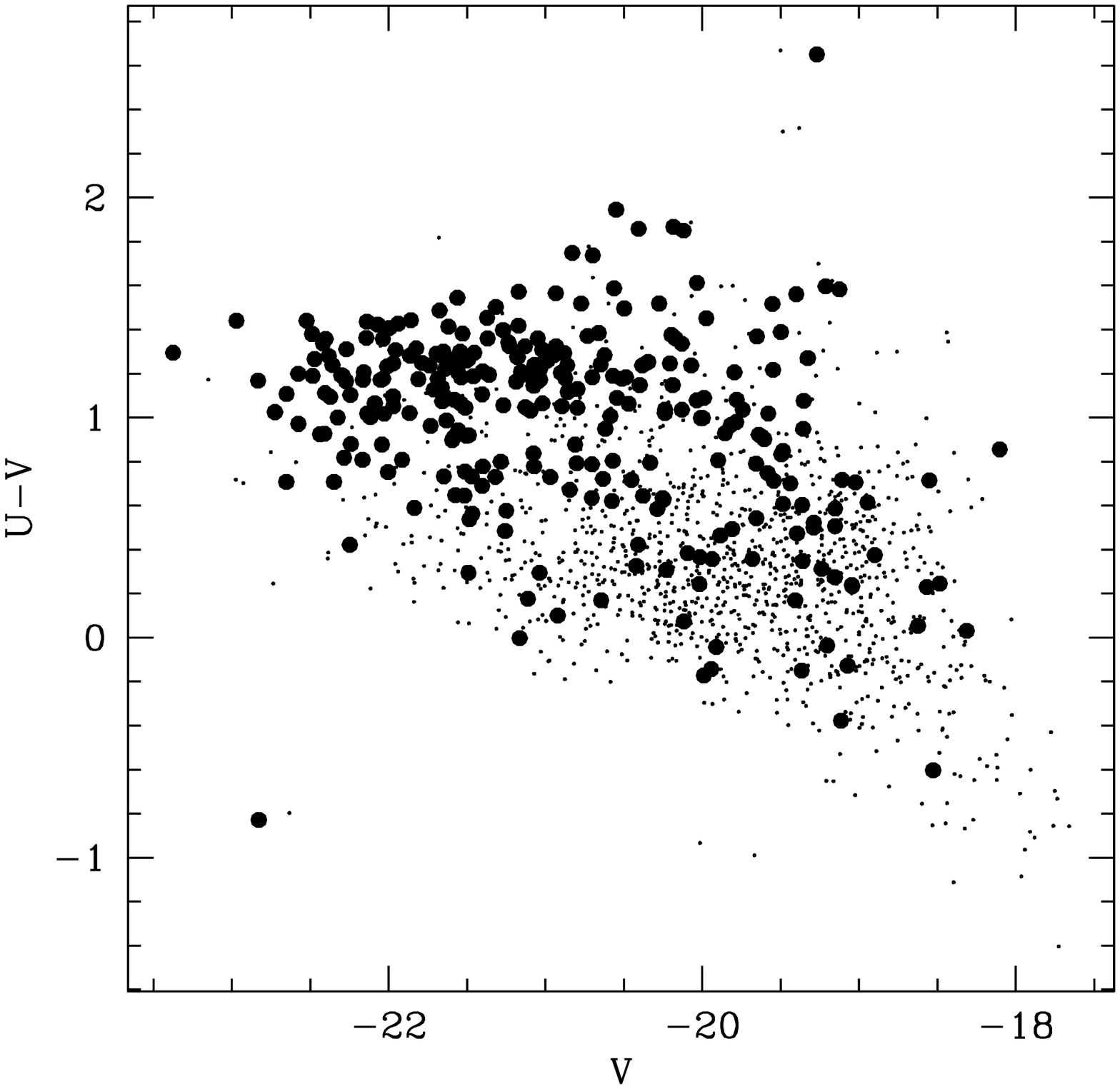}}
\caption{\footnotesize
{\bf a)} U-V rest-frame color distribution; the shaded histogram is
for the LT population, while the heavy line histogram is for the ET
population. The inset shows the same histograms drawn only for the
higher S/N objects.  
{\bf b)} U-V vs V color-magnitude relation; filled dots are ET
objects, tiny dots are LT ones. Both plots include only the objects
with redshift within the interval 0.6-0.8 from the ``spectroscopic
sub-sample''  }
\label{hist}
\end{center}
\end{figure*}

To a first approximation the definition of the separating boundary
between LT and ET galaxies in the D4000 vs. EW[OII] parameters space could be
a vertical line at EW[OII]=10\AA~, the approximate detection limit for the
[OII] line in our data. This definition, however, would lead us to include
many objects with very low D4000 and a real, albeit undetected, [OII] emission
in the ET sample, which is obviously contrary to the natural definition of
early-type galaxies as objects with an evolved stellar population 
\citep[large D4000; see][]{bimodal_4000}
and no current star formation activity (no [OII] emission line). 
To alleviate this problem, we have checked various definitions of the ET-LT 
separating boundary line that would keep constant the ratio of ET to LT galaxies.
We have built a composite spectrum of the ET population for each one of
the definitions so that the higher signal-to-noise ratio thus achieved would 
allow a robust measurement of the [OII] emission at fainter intensities.
 Then we have selected the subdivision that minimizes the equivalent width of the 
[OII] line in the composite spectrum of the resulting ET population. This 
optimization procedure results in a slightly tilted boundary line described 
by the relation: 
\\\\
$D_n(4000) + EW([OII])/15.0 = 0.7$ 
\\\\ 
According to this definition, ET galaxies have $D_n(4000) + EW([OII])/15.0$ 
above 0.7, while LT galaxies are
below that value. Purely for comparison, in figure~\ref{d4oii} we have also
plotted as large numbered circles the $D_n(4000)$ and $EW([OII])$ values
measured on Virgo cluster templates from \citet{gavazzi_cri}.
Table~\ref{tab_virgo} recaps the legend for the circles. It should not be
surprising that our ET-LT boundary is including early spirals within the ET
population, and only Sc and later types within the LT population. This is a
rather general property of classification schemes based on galaxy color or
spectral properties. For example, the earliest spiral galaxy SED presented by
\citet{cww} is that typical of Sbc galaxies, and this same SED was used by
\citet{cfrs_iii} to separate red and blue galaxies in their CFRS galaxy
sample. Similarly, the earliest spiral galaxy color-color track used by
\citet{adelberger} in defining the color selection criteria for isolating
star-forming galaxies in the redshift range $1<z<3$ is that typical of an Sb
galaxy.

We consider this spectral classification as a better substitute for a
true morphological classification with respect to color for a number of
reasons: it is based on directly measurable quantities and avoids the
uncertainties involved in estimating rest-frame colors; it is directly
based on two important physical properties like the age of the stellar 
population and the amount of star formation activity taking place in each galaxy;
unlike colors, it is minimally affected by the unknown amount of
reddening inside each galaxy. One possible limitation affecting our
spectral classification is the fact that a number of objects with
relatively young stellar population, as witnessed by the small value
of their D4000, is included within the ET population. However we
prefer not to exclude these objects by some modification of the ET-LT
separation boundary, because any such a modification would make our
classification scheme much more vulnerable to progenitor bias
effects \citep{progenitor_bias}. 
With the current scheme as soon as a galaxy has completed the
bulk of its star formation activity and is starting the purely passive
evolution phase it becomes an ET object, and it remains such at all
subsequent times, as its stellar population ages and the D4000
amplitude in its spectrum increases. Instead, any significant star
formation activity would move an object towards the left in the
figure~\ref{d4oii} diagram, effectively removing it from the ET
population. This minimization of progenitor bias effects on our
classification scheme is the main reason we can use a
redshift-independent classification scheme in our analysis.

\subsection{The contamination effect}
\label{sec:contamination}

Having obtained an early- vs. late-type galaxy separation based on
spectral properties, we analyse what is the color distribution of
these two categories, and compare this spectral classification with
the ``red-peak'' color one. Figure~\ref{hist}a shows the global U-V
rest-frame color distribution for our ``spectroscopic sub-sample'' 
in the redshift interval where the peak of the N(z) distribution is 
located \citep[$0.6<z<0.8$, see][]{vvds_main}. In this paragraph we
focus our analysis on this redshift bin before extending it to the
whole redshift range to study the evolution of the contamination.
 
The shaded histogram represents the color distribution for the LT 
objects, while the thick line shows the distribution for the ET 
population. The thin line histogram represents the sum of the two 
distributions. As expected, the red
peak of the global color distribution is mainly populated by ET
objects, while the blue peak is mainly populated by LT ones, but a
rather strong contamination between the two populations is present.
The inset plotted in figure~\ref{hist}a shows the
same histograms drawn considering only the objects with higher S/N
spectra (2229 objects); it is easy to see how very little difference 
is observed when removing low S/N spectra objects, for which spectral 
feature measurements are more uncertain. We therefore conclude that
uncertainties in the spectral classification are not the primary
source for the significant overlap in color space between the two
populations.
The color distributions for the two populations have a very
significant overlap at intermediate colors, and a simple color cut is
not able to separate the galaxy populations according to the spectral
classification defined above. 
In other words, a purely color based classification does not completely
account for the different stellar populations and star formation
history properties.

\begin{figure}[t!]
\begin{center}
\resizebox{8.5truecm}{!}{\includegraphics[clip=true]{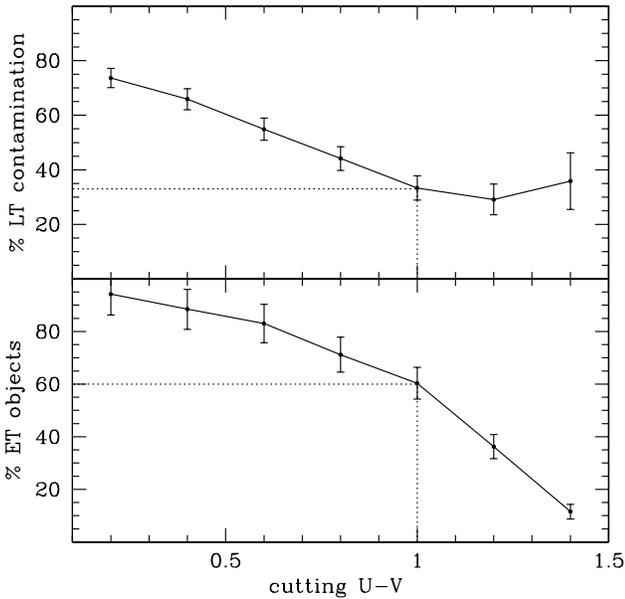}}
\caption{\footnotesize
The ``contamination'' effect in the redshift range $0.6<z<0.8$.  The
fraction of the whole ET population which is included in the resulting
RED population (lower panel) and the ``contamination'' of the
resulting RED sample (upper panel) as a function of the limiting color
used to define it.  }
\label{contcol}
\end{center}
\end{figure}

Similar results have been obtained by \citet{mass_func_deep} using the
DEEP2 data (see their figure 1), although they use a different color
and a slightly different criterion for their spectral classification
of passive and star-forming galaxies.  The same contamination effect
is visible in the color-magnitude relation plot
(figure~\ref{hist}b).  Projecting the color-magnitude relation along
the slope of the red sequence to obtain ``corrected'' color
histograms, as done for example in B04, does not significantly change
the situation.

From the total histogram in figure~\ref{hist} we see that a purely
color-based definition of the red peak population (hereafter RED 
population) would be that of galaxies with $U-V>1.0$, in agreement 
with the color cut implemented by B04. This is the color where a 
local minimum is observed in the color distribution, and it is also 
the color at which we observe the transition from an LT dominated 
population to an ET dominated one. However, because of the significant 
color overlap between the two populations, it is clear that the 
resulting RED population includes not only ET (i.e. really passive) 
objects, but also a significant fraction of LT (i.e. star-forming) 
ones. Moreover this RED population does not include a significant 
fraction of ET objects which have bluer colors.

In figure~\ref{contcol} we quantify such ``contamination''.  
The lower panel shows, as a function of the color cut we use to 
define the RED population, the fraction of the whole ET population 
which is included in it.  The upper panel shows, on the contrary, 
the ``contamination'' of the resulting RED sample, i.e. which 
percentage of the total RED population is actually composed of LT objects.

Error bars in this and the following two figures have been computed
using simple Montecarlo simulations. The main sources of uncertainty,
beside the Poissonian error on population counts, are the
uncertainties on the spectroscopic and photometric measurements. In
our simulations we generated mock datasets by adding to the measured
values random offsets, generated from a Gaussian distribution
characterized by a $\sigma$ equal to the measurement uncertainty for
the given quantity. Then we repeated our determinations of the
contamination effect for all the mock datasets to obtain an estimate
for standard deviation of each data point in the plots.

To these Montecarlo uncertainty estimates we have also added the
appropriate Poissonian counting uncertainties. The small values for
the contamination uncertainties demonstrate that the effect we are
witnessing is real, and not the result of some bias introduced by
measurement uncertainties. Using the $U-V>1.0$ cut we include in the
RED population only $\sim 60\%$ of the ET objects, the remaining $\sim
40\%$ having bluer colors. At the same time, 35\% of our RED
population will be actually constituted by LT objects which have red
colors, but show clear spectroscopic indication of star formation
activity.

To summarize, 65\% of our RED population are ET population (and
these objects are 60\% of all the ET objects) while 35\% are LT
objects.

As shown in section \ref{sec:redbias} a small fraction of photometric 
early-type objects is missing in the spectroscopic sample due to the lower 
efficiency in measuring the redshift for these galaxies with respect to 
the late-type ones. However, this fraction is small and it does not change 
the reliability of our results; adding a few percents of objects in the red 
peak would not change significantly the observed contamination, even if all 
the ``new'' red objects were to be ET objects. We therefore consider the bias 
introduced by the lower redshift measurement efficiency to be almost negligible.

\subsection{Contamination evolution with redshift}

\begin{figure}[t!]
\resizebox{\hsize}{!}{\includegraphics[clip=true]{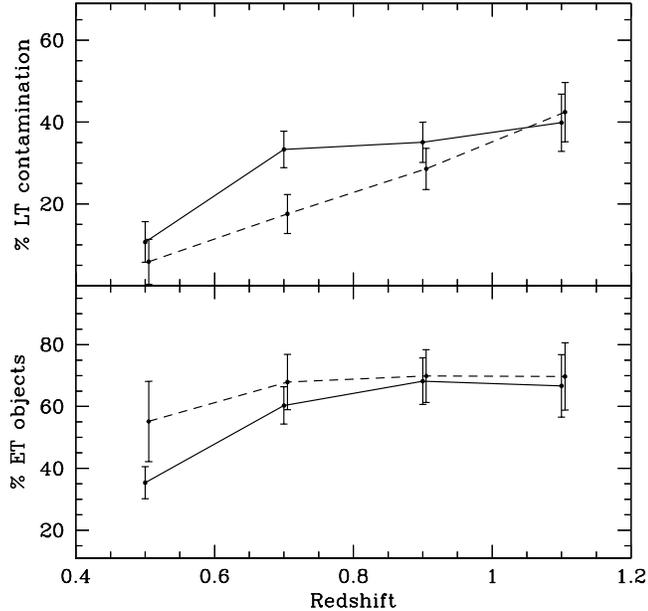}}
\caption{\footnotesize
The fraction of the ET objects which are included in the RED
population and the corresponding LT contamination computed for a
limiting color $U-V=1.0$ as a function of redshift. The solid line 
is for the whole ``spectroscopic sub-sample''; the dashed line is
computed using only objects brighter than $M_V= -20$. 
}
\label{popsz_fc}
\end{figure}

The contamination effect changes somewhat with redshift as 
shown in figure~\ref{popsz_fc} (solid line). The definition of the RED
population at different redshifts is derived from the color distributions
shown in figure \ref{colmag}, where a local minimum in the distributions 
is located at $\sim U-V=1.0$ for all redshift bins (within the small uncertainty
introduced by the arbitrary binning of the data). We therefore define as RED
galaxies all objects with color $U-V>1.0$ within all our redshift bins.

The fraction of RED ET objects shows little evolution over most 
of the redshift range we cover with our sample, while the RED objects 
LT contamination becomes more important with redshift as 
could be expected because of the well known widespread increase in star 
formation activity from the local universe to redshift 1-2 that should
results in a higher contaminating fraction of dusty starburst galaxies
in the RED population. 

One exception is represented by the low fraction of ET objects observed in
the lowermost redshift bin, but this result is significantly affected by 
our sample characteristics.
The first effect is due to the VVDS relatively small field of view;
the number of bright objects we have in the lowest redshift
bins is small, and since these bright objects are predominantly 
red galaxies, this results in a smaller fraction of ET galaxies 
included in the red sample.
Moreover in the lowest redshift bins we include faint objects which
are not observable at higher redshift. As already discussed by
\citet{early_progenitors}, we observe that the type mix in the RED 
population depends on the range of galaxy luminosities that are included 
in the sample.

By restricting ourselves only to galaxies brighter than $M_V=-20$ (dashed
line in figure~\ref{popsz_fc}) we observe a much smaller evolution with
redshift of the RED ET fraction, and a somewhat smaller contamination 
from LT objects. Although some of these differences could be
due to larger errors in the measurements of photometric and
spectroscopic parameters for the fainter objects, we clearly observe
that transition objects (blue ET or red LT galaxies) are found
preferentially at fainter luminosities.

\begin{figure}[t!]
\resizebox{\hsize}{!}{\includegraphics[clip=true]{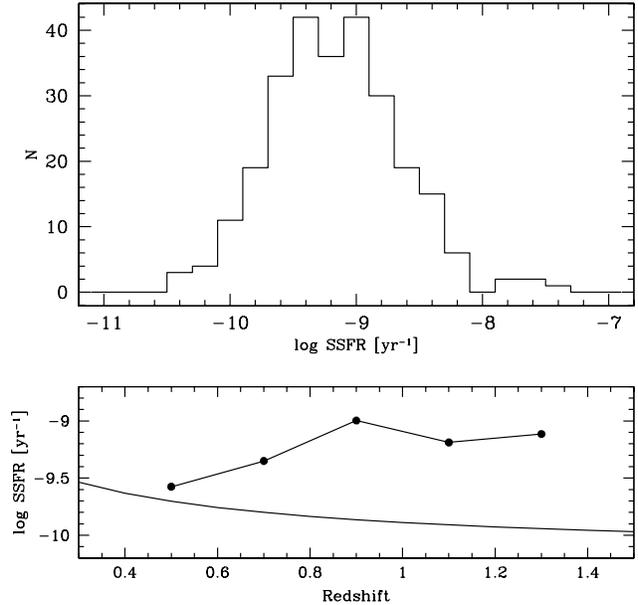}}
\caption{\footnotesize
In the upper panel the distribution of star formation rates for the
contaminating LT population among the RED galaxies is shown.  The
evolution of the median value with redshift is plotted in the lower
panel. The solid line is the SSFR which is required to double the
galaxy stellar mass between the given redshift and the present time,
which discriminates between quiescent and strongly star-forming
galaxies.  }
\label{sfr}
\end{figure}

To quantify how different from a passively evolving object is the typical
contaminant galaxy, we have derived current specific star formation
rate estimates for all the RED population LT galaxies, using the [OII]
luminosity conversion factor given by \citet{kenni_sfr}. The resulting
distribution of specific star formation rates is shown in the upper
panel of figure~\ref{sfr} and has a median value of $6.76 \times 10^{-10}~
yr^{-1}$. The lower panel of figure~\ref{sfr} shows the evolution of
the median value with redshift. The solid line in this plot represents the 
SSFR which is required to double the galaxy stellar mass between the given
redshift and the present time, assuming a constant SFR; this
``doubling'' SSFR is used to discriminate between quiescent and
star-forming galaxies. It is easy to see that, at all redshifts, the
median SSFR value is above the threshold. It is clear that a
significant number of dusty starburst galaxies is included in the
RED population, and its presence should be accounted for in studying
the evolution of the RED population properties. This result is in
agreement with the findings of \citet{giallongo_dusty}, who found some
35\% of dusty starburst objects within the red population of their HDF
N+S sample, and with those of \citet{paoloc_morph}, who found a
similar fraction of 33\% edge-on spirals within their sample of red 
galaxies in the COSMOS survey field.
We also notice that the luminosity dependence of the observed
LT contamination brings our results in agreement with the findings of
\citet{redpop_gems} who measured a contamination of some 15\% from 
morphologically classified late-type galaxies in their bright red galaxies
sample. In a similar redshift bin,
using a similar sample (our bright sample limited to $M_V<-20$), we measure
a contamination from LT objects which is between 13 and 21 percent 
(see figure \ref{popsz_fc}, dashed line).

\begin{figure}[t!]
\resizebox{\hsize}{!}{\includegraphics[clip=true]{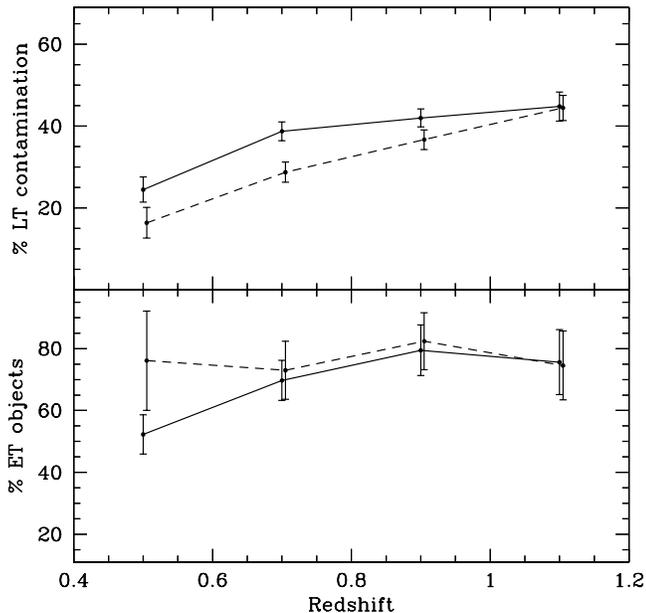}}
\caption{\footnotesize
The fraction of the ET objects which are included in the RED
population and the corresponding LT contamination computed for
multi-color types (see section \ref{multicolor}). The solid line 
is for the whole ``spectroscopic sub-sample''; the dashed line is
computed using only objects brighter than $M_V= -20$. }
\label{photo_type}
\end{figure}

\subsection{The multi-color type approach}
\label{multicolor}

Given the not entirely satisfactory results obtained by a pure color
selection in isolating an old, passively evolving galaxy population,
we have considered also an alternative approach towards this goal,
still based solely on photometric data, using the ``photometric type''
classification scheme described in section \ref{sec:redbias}.

This approach does not account for any global SED variation over cosmic
time as a result of the evolution of the stellar populations, nor for
the effects of a progenitor bias \citep{progenitor_bias} against the
early-type galaxy population. Still, it provides a very simple and
model-independent ``no evolution'' reference for any other, more
complex modeling of galaxy populations evolution. 

Figure~\ref{photo_type} shows the fraction of the ET objects which 
are included in the ``photometric early-type population'' and the 
corresponding LT contamination computed for multi-color types.
Comparing figure~\ref{popsz_fc} with figure~\ref{photo_type} we see
that, the multi-color type approach has a somewhat higher efficiency 
in separating old passive galaxies from star forming ones. Indeed with 
the single $U-V$ color separation we isolate on average (over the redshift 
range covered by our sample) about 60\% of the ET galaxies, whereas in 
the case of the multi-color separation this fraction is $\sim$ 70\%. 
We note that the 35\% level of LT contamination is similar in both cases.  
Also the trends with redshift of both the fractions of selected ET 
galaxies and of LT galaxies contamination are comparable to the single 
color approach, since in both cases the selection criterion does not 
evolve with redshift.

Given the above results, we conclude that the photometric types
approach is to be preferred to the color selection one, as it provides
a finer degree of subdivision between galaxy properties than the color
separation, making it a better tool for the study of galaxy
evolution. Again, like in the case of the color selection, we observe
a luminosity dependence for the population mix; limiting our sample
to the bright $M_B<-20$ objects we observe a smaller evolution 
with redshift of the RED ET fraction, and a somewhat smaller 
contamination from LT objects.

\section{The CMR of the spectral early-type population}
\label{sec:cmr_early}

\begin{figure}[t!]
\resizebox{\hsize}{!}{\includegraphics[clip=true]{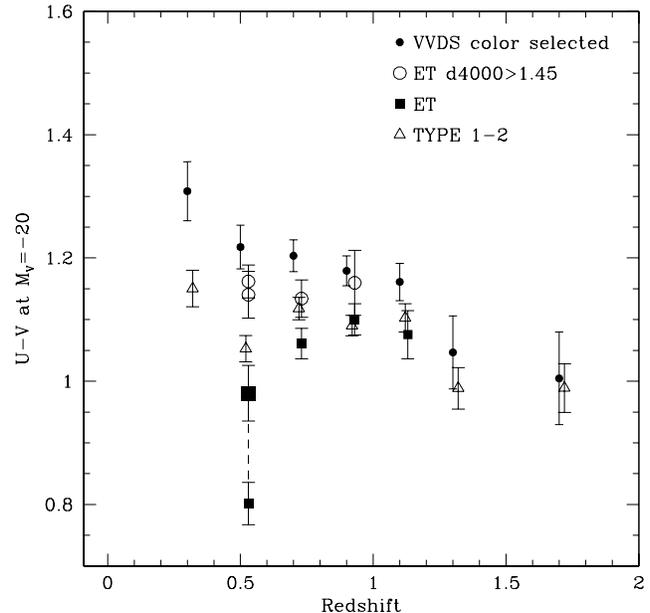}}
\caption{\footnotesize
Color evolution for four classes of objects as measured by the value
of the intercept of the CMR at $M_V=-20$. Small filled circles are the
VVDS color selected data (as in figure \ref{red_evol_1}); open
triangles are the photometric early-type population; filled squares
are the ET population; open circles are the ``oldest'' part of the ET
population, i.e.  the ET objects for which the value of the
D$_n$(4000) is higher than 1.45.  
The lowermost redshift point for the spectroscopical ET sample
is plotted also considering only objects brighter than $M_V=-20$
(big filled square)
}
\label{red_evol_2}
\end{figure}

In section~\ref{sec:cmr_red} we have analyzed the properties of the
CMR of color-selected early-type galaxies in our sample, showing that
there is a generally good agreement between the observed evolution of
the CMR and the expectations for a purely passively evolving
population. Having shown that a very significant contamination of 
the red population from (presumably dust-reddened) actively 
star-forming galaxies is present, we must consider the possibility
that the above agreement is in fact a rather fortuitous one.

In figure~\ref{red_evol_2} we now plot the evolution of the CMR for
the spectroscopically selected ET galaxies (filled squares), and also
for the multi-color selected early-type ones (open triangles). 
Notice how the lowermost redshift point for the spectroscopically
defined ET sample is abnormally blue, in agreement with the small 
fraction of RED ET galaxies shown in figure~\ref{popsz_fc} for the 
same redshift bin. 
If we consider once again only objects brighter than $M_V=-20$,
the point moves to a significantly redder color (as shown in the figure
by the big filled square).  

From a comparison with figure~\ref{red_evol_1} we can see that only small
differences are present in the CMR of the different early-type samples
(the solid circles in this figure are the same as in
figure~\ref{red_evol_1}, representing the evolution of the CMR for the
color-selected red galaxies). The color-magnitude relations of ET and
multi-color selected early-type galaxies are on average 0.15
magnitudes bluer than that of red galaxies, as could be expected from
the presence of a blue color tail in these galaxy samples. 
Even restricting the ET sample to the ``oldest'' part of the ET 
population, i.e. the ET objects for which the value of the D$_n$(4000) 
is higher than 1.45 (open circles in figure~\ref{red_evol_2}) does 
not change very significantly the situation, although the CMR for 
these objects is slightly redder than that of the full ET sample.

The single-color, multi-color, and spectroscopically selected samples
of early-type galaxies we have obtained are not independent, having
some 60\% of galaxies in common, and therefore a certain degree of
similarity in the various color-magnitude relations we have derived is
expected. But we see from figure~\ref{red_evol_2} how the red
sample, including some 40\% contamination from actively star-forming
galaxies, and the subset of the ``oldest'' ET galaxies, composed
almost exclusively by galaxies with an old stellar population and no 
current star-formation activity, are described by essentially the 
same CMR, over a rather large redshift interval. These findings 
support the conclusion that the average color of the red objects, as 
measured by the CMR value at fixed luminosity, may not be a very 
sensitive indicator for measuring the evolution of the early-type 
galaxy population.

This important issue deserves a more extended analysis. In future 
papers we will study the spectral classification in more detail 
and expand our analysis on the effects that different sample 
selection criteria can have on the inferred properties of the early- 
and late-type galaxy populations.

\section{Summary and conclusions}

In this paper we have conducted the study of the rest-frame color
distributions on a sample of 6291 galaxies from the first epoch VVDS.

We find that the well-studied local bimodal distribution is still
present at previously unexplored redshifts up to at least z=1.5
(the mean redshift for the galaxies in the redshift bin $1.2<z<2.0$).

Given the large interval of redshift and luminosity spanned by our
sample, we were able to study also the dependence of rest-frame colors
distributions on these quantities.  We have shown how, at each
redshift, faint galaxies are bluer than the luminous ones.  Moreover,
at any given luminosity, galaxies become bluer as redshift increases
and the amount of this effect is quite independent on the
luminosity. This result extends for the first time the findings of
previous redshift surveys significantly over the $z\approx1$ limit.
We have extended up to $z\sim2$ the results of B04 about the evolution
of the CMR of red galaxies, finding a slightly smaller evolution and a
better agreement with the $z=0$ point.

We have analyzed the galaxy population which constitutes the red-peak
of the color distributions finding that a significant fraction is
composed of star forming objects. The detailed quantification of this
effect at various redshifts proves how a simple color selection is not
able to reliably isolate early-type objects. The different approach
used within the VVDS to isolate galaxy types, based solely on
photometric data and local spectral templates fitting, has been shown
to have a slightly better capability of isolating passive object than
that based on a single color.

Using a robust spectral classification of early-type galaxies in the
D$_n$(4000)/EW([OII]) plane, we have demonstrated that a population of
galaxies selected only on the basis of their red colors is not only
including early-type galaxies with little star formation, but also a
significant contamination by star-forming galaxies. We conclude that
there is no one to one correspondence between red-sequence and 
"dead and old" galaxies, and that selecting galaxies only on the 
basis of their colors can be misleading in estimating the evolution 
of old and passively evolving galaxies.

\begin{acknowledgements}
This research has been developed within the framework of the VVDS
consortium.\\
This work has been partially supported by the
CNRS-INSU and its Programme National de Cosmologie (France),
and by Italian Ministry (MIUR) grants
COFIN2000 (MM02037133) and COFIN2003 (num.2003020150).\\
The VLT-VIMOS observations have been carried out on guaranteed
time (GTO) allocated by the European Southern Observatory (ESO)
to the VIRMOS consortium, under a contractual agreement between the
Centre National de la Recherche Scientifique of France, heading
a consortium of French and Italian institutes, and ESO,
to design, manufacture and test the VIMOS instrument.\\
DV acknowledges supports by the European Commission through a 
Marie Curie ERG grant (No. MERG-CT-2005-021704).\\
We thank the anonymous referee for many valuable suggestions that
helped improving the overall quality of this paper.

\end{acknowledgements}

%\begin{thebibliography}
%\bibliographystyle{aa}
\bibliography{5942}     

\begin{thebibliography}{60}
\expandafter\ifx\csname natexlab\endcsname\relax\def\natexlab#1{#1}\fi

\bibitem[{{Adelberger} {et~al.}(2004){Adelberger}, {Steidel}, {Shapley},
  {Hunt}, {Erb}, {Reddy}, \& {Pettini}}]{adelberger}
{Adelberger}, K.~L., {Steidel}, C.~C., {Shapley}, A.~E., {et~al.} 2004, \apj,
  607, 226

\bibitem[{{Baldry} {et~al.}(2004{\natexlab{a}}){Baldry}, {Balogh}, {Bower},
  {Glazebrook}, \& {Nichol}}]{bimodal_8}
{Baldry}, I.~K., {Balogh}, M.~L., {Bower}, R., {Glazebrook}, K., \& {Nichol},
  R.~C. 2004{\natexlab{a}}, in AIP Conf. Proc. 743: The New Cosmology:
  Conference on Strings and Cosmology, 106--119

\bibitem[{{Baldry} {et~al.}(2004{\natexlab{b}}){Baldry}, {Glazebrook},
  {Brinkmann}, {Ivezi{\' c}}, {Lupton}, {Nichol}, \& {Szalay}}]{bimodal_9}
{Baldry}, I.~K., {Glazebrook}, K., {Brinkmann}, J., {et~al.}
  2004{\natexlab{b}}, \apj, 600, 681

\bibitem[{{Balogh} {et~al.}(2004){Balogh}, {Eke}, {Miller}, {Lewis}, {Bower},
  {Couch}, {Nichol}, {Bland-Hawthorn}, {Baldry}, {Baugh}, {Bridges}, {Cannon},
  {Cole}, {Colless}, {Collins}, {Cross}, {Dalton}, {de Propris}, {Driver},
  {Efstathiou}, {Ellis}, {Frenk}, {Glazebrook}, {Gomez}, {Gray}, {Hawkins},
  {Jackson}, {Lahav}, {Lumsden}, {Maddox}, {Madgwick}, {Norberg}, {Peacock},
  {Percival}, {Peterson}, {Sutherland}, \& {Taylor}}]{bimodal_ha}
{Balogh}, M., {Eke}, V., {Miller}, C., {et~al.} 2004, \mnras, 348, 1355

\bibitem[{{Balogh} {et~al.}(1999){Balogh}, {Morris}, {Yee}, {Carlberg}, \&
  {Ellingson}}]{balogh_spectral_indexes}
{Balogh}, M.~L., {Morris}, S.~L., {Yee}, H.~K.~C., {Carlberg}, R.~G., \&
  {Ellingson}, E. 1999, \apj, 527, 54

\bibitem[{{Beers} {et~al.}(1990){Beers}, {Flynn}, \& {Gebhardt}}]{biweight}
{Beers}, T.~C., {Flynn}, K., \& {Gebhardt}, K. 1990, \aj, 100, 32

\bibitem[{{Bell} {et~al.}(2004{\natexlab{a}}){Bell}, {McIntosh}, {Barden},
  {Wolf}, {Caldwell}, {Rix}, {Beckwith}, {Borch}, {H{\"a}ussler}, {Jahnke},
  {Jogee}, {Meisenheimer}, {Peng}, {Sanchez}, {Somerville}, \&
  {Wisotzki}}]{redpop_gems}
{Bell}, E.~F., {McIntosh}, D.~H., {Barden}, M., {et~al.} 2004{\natexlab{a}},
  \apjl, 600, L11

\bibitem[{{Bell} {et~al.}(2004{\natexlab{b}}){Bell}, {Wolf}, {Meisenheimer},
  {Rix}, {Borch}, {Dye}, {Kleinheinrich}, {Wisotzki}, \&
  {McIntosh}}]{bimodal_bell}
{Bell}, E.~F., {Wolf}, C., {Meisenheimer}, K., {et~al.} 2004{\natexlab{b}},
  \apj, 608, 752

\bibitem[{{Bernardi} {et~al.}(2003){Bernardi}, {Sheth}, {Annis}, {Burles},
  {Finkbeiner}, {Lupton}, {Schlegel}, {SubbaRao}, {Bahcall}, {Blakeslee},
  {Brinkmann}, {Castander}, {Connolly}, {Csabai}, {Doi}, {Fukugita}, {Frieman},
  {Heckman}, {Hennessy}, {Ivezi{\'c}}, {Knapp}, {Lamb}, {McKay}, {Munn},
  {Nichol}, {Okamura}, {Schneider}, {Thakar}, \& {York}}]{bernardi03}
{Bernardi}, M., {Sheth}, R.~K., {Annis}, J., {et~al.} 2003, \aj, 125, 1882

\bibitem[{{Bottini} {et~al.}(2005){Bottini}, {Garilli}, {Maccagni}, {Tresse},
  {Le Brun}, {Le F{\`e}vre}, {Picat}, {Scaramella}, {Scodeggio}, {Vettolani},
  {Zanichelli}, {Adami}, {Arnaboldi}, {Arnouts}, {Bardelli}, {Bolzonella},
  {Cappi}, {Charlot}, {Ciliegi}, {Contini}, {Foucaud}, {Franzetti}, {Guzzo},
  {Ilbert}, {Iovino}, {McCracken}, {Marano}, {Marinoni}, {Mathez}, {Mazure},
  {Meneux}, {Merighi}, {Paltani}, {Pollo}, {Pozzetti}, {Radovich}, {Zamorani},
  \& {Zucca}}]{vmmps}
{Bottini}, D., {Garilli}, B., {Maccagni}, D., {et~al.} 2005, \pasp, 117, 996

\bibitem[{{Bower} {et~al.}(1992){Bower}, {Lucey}, \& {Ellis}}]{local_cmr}
{Bower}, R.~G., {Lucey}, J.~R., \& {Ellis}, R.~S. 1992, \mnras, 254, 601

\bibitem[{{Brinchmann} {et~al.}(2004){Brinchmann}, {Charlot}, {White},
  {Tremonti}, {Kauffmann}, {Heckman}, \& {Brinkmann}}]{bimodal_sfh}
{Brinchmann}, J., {Charlot}, S., {White}, S.~D.~M., {et~al.} 2004, \mnras, 351,
  1151

\bibitem[{{Bruzual}(1983)}]{bruzual83}
{Bruzual}, G. 1983, \apj, 273, 105

\bibitem[{{Bruzual} \& {Charlot}(2003)}]{bc_last}
{Bruzual}, G. \& {Charlot}, S. 2003, \mnras, 344, 1000

\bibitem[{{Budav{\' a}ri} {et~al.}(2003){Budav{\' a}ri}, {Connolly}, {Szalay},
  {Szapudi}, {Csabai}, {Scranton}, {Bahcall}, {Brinkmann}, {Eisenstein},
  {Frieman}, {Fukugita}, {Gunn}, {Johnston}, {Kent}, {Loveday}, {Lupton},
  {Tegmark}, {Thakar}, {Yanny}, {York}, \& {Zehavi}}]{bimodal_clustering}
{Budav{\' a}ri}, T., {Connolly}, A.~J., {Szalay}, A.~S., {et~al.} 2003, \apj,
  595, 59

\bibitem[{{Bundy} {et~al.}(2005){Bundy}, {Ellis}, {Conselice}, {Cooper},
  {Weiner}, {Taylor}, {Willmer}, \& {DEEP2 Team}}]{mass_func_deep}
{Bundy}, K., {Ellis}, R.~S., {Conselice}, C.~J., {et~al.} 2005, American
  Astronomical Society Meeting Abstracts, 207,

\bibitem[{{Cassata} {et~al.}(2007){Cassata}, {Guzzo}, {Franceschini},
  {Scoville}, {Capak}, {Ellis}, \& {Koekemoer}}]{paoloc_morph}
{Cassata}, P., {Guzzo}, L., {Franceschini}, A., {et~al.} 2007, astro-ph/0701483

\bibitem[{{Cimatti} {et~al.}(2002){Cimatti}, {Daddi}, {Mignoli}, {Pozzetti},
  {Renzini}, {Zamorani}, {Broadhurst}, {Fontana}, {Saracco}, {Poli},
  {Cristiani}, {D'Odorico}, {Giallongo}, {Gilmozzi}, \& {Menci}}]{cimatti_eros}
{Cimatti}, A., {Daddi}, E., {Mignoli}, M., {et~al.} 2002, \aap, 381, L68

\bibitem[{{Coleman} {et~al.}(1980){Coleman}, {Wu}, \& {Weedman}}]{cww}
{Coleman}, G.~D., {Wu}, C.-C., \& {Weedman}, D.~W. 1980, \apjs, 43, 393

\bibitem[{{Cowie} {et~al.}(1996){Cowie}, {Songaila}, {Hu}, \& {Cohen}}]{hdfs}
{Cowie}, L.~L., {Songaila}, A., {Hu}, E.~M., \& {Cohen}, J.~G. 1996, \aj, 112,
  839

\bibitem[{{Cucciati} {et~al.}(2006){Cucciati}, {Iovino}, {Marinoni}, {Ilbert},
  {Bardelli}, {Franzetti}, {Le F{\`e}vre}, {Pollo}, {Zamorani}, {Cappi},
  {Guzzo}, {McCracken}, {Meneux}, {Scaramella}, {Scodeggio}, {Tresse}, {Zucca},
  {Bottini}, {Garilli}, {Le Brun}, {Maccagni}, {Picat}, {Vettolani},
  {Zanichelli}, {Adami}, {Arnaboldi}, {Arnouts}, {Bolzonella}, {Charlot},
  {Ciliegi}, {Contini}, {Foucaud}, {Gavignaud}, {Marano}, {Mazure}, {Merighi},
  {Paltani}, {Pell{\`o}}, {Pozzetti}, {Radovich}, {Bondi}, {Bongiorno},
  {Busarello}, {de La Torre}, {Gregorini}, {Lamareille}, {Mathez}, {Mellier},
  {Merluzzi}, {Ripepi}, {Rizzo}, {Temporin}, \& {Vergani}}]{vvds_density}
{Cucciati}, O., {Iovino}, A., {Marinoni}, C., {et~al.} 2006, \aap, 458, 39

\bibitem[{{Dekel} \& {Birnboim}(2006)}]{bimo_models_2}
{Dekel}, A. \& {Birnboim}, Y. 2006, \mnras, 368, 2

\bibitem[{{Dressler}(1980)}]{dens_morph}
{Dressler}, A. 1980, \apj, 236, 351

\bibitem[{{Gavazzi} {et~al.}(2002){Gavazzi}, {Bonfanti}, {Sanvito}, {Boselli},
  \& {Scodeggio}}]{gavazzi_cri}
{Gavazzi}, G., {Bonfanti}, C., {Sanvito}, G., {Boselli}, A., \& {Scodeggio}, M.
  2002, \apj, 576, 135

\bibitem[{{Gavazzi} {et~al.}(2003){Gavazzi}, {Boselli}, {Donati}, {Franzetti},
  \& {Scodeggio}}]{goldmine}
{Gavazzi}, G., {Boselli}, A., {Donati}, A., {Franzetti}, P., \& {Scodeggio}, M.
  2003, \aap, 400, 451

\bibitem[{{Gavazzi} {et~al.}(1996){Gavazzi}, {Pierini}, \&
  {Boselli}}]{gavazzi_mass}
{Gavazzi}, G., {Pierini}, D., \& {Boselli}, A. 1996, \aap, 312, 397

\bibitem[{{Giallongo} {et~al.}(2005){Giallongo}, {Salimbeni}, {Menci},
  {Zamorani}, {Fontana}, {Dickinson}, {Cristiani}, \&
  {Pozzetti}}]{giallongo_dusty}
{Giallongo}, E., {Salimbeni}, S., {Menci}, N., {et~al.} 2005, \apj, 622, 116

\bibitem[{{Ilbert} {et~al.}(2006){Ilbert}, {Arnouts}, {McCracken},
  {Bolzonella}, {Bertin}, {Le F{\`e}vre}, {Mellier}, {Zamorani}, {Pell{\`o}},
  {Iovino}, {Tresse}, {Le Brun}, {Bottini}, {Garilli}, {Maccagni}, {Picat},
  {Scaramella}, {Scodeggio}, {Vettolani}, {Zanichelli}, {Adami}, {Bardelli},
  {Cappi}, {Charlot}, {Ciliegi}, {Contini}, {Cucciati}, {Foucaud}, {Franzetti},
  {Gavignaud}, {Guzzo}, {Marano}, {Marinoni}, {Mazure}, {Meneux}, {Merighi},
  {Paltani}, {Pollo}, {Pozzetti}, {Radovich}, {Zucca}, {Bondi}, {Bongiorno},
  {Busarello}, {de La Torre}, {Gregorini}, {Lamareille}, {Mathez}, {Merluzzi},
  {Ripepi}, {Rizzo}, \& {Vergani}}]{photoz_cfhtls}
{Ilbert}, O., {Arnouts}, S., {McCracken}, H.~J., {et~al.} 2006, \aap, 457, 841

\bibitem[{{Ilbert} {et~al.}(2004){Ilbert}, {Tresse}, {Arnouts}, {Zucca},
  {Bardelli}, {Zamorani}, {Adami}, {Cappi}, {Garilli}, {Le F{\`e}vre},
  {Maccagni}, {Meneux}, {Scaramella}, {Scodeggio}, {Vettolani}, \&
  {Zanichelli}}]{lumfunc_bias}
{Ilbert}, O., {Tresse}, L., {Arnouts}, S., {et~al.} 2004, \mnras, 351, 541

\bibitem[{{Ilbert} {et~al.}(2005){Ilbert}, {Tresse}, {Zucca}, {Bardelli},
  {Arnouts}, {Zamorani}, {Pozzetti}, {Bottini}, {Garilli}, {Le Brun}, {Le
  F{\`e}vre}, {Maccagni}, {Picat}, {Scaramella}, {Scodeggio}, {Vettolani},
  {Zanichelli}, {Adami}, {Arnaboldi}, {Bolzonella}, {Cappi}, {Charlot},
  {Contini}, {Foucaud}, {Franzetti}, {Gavignaud}, {Guzzo}, {Iovino},
  {McCracken}, {Marano}, {Marinoni}, {Mathez}, {Mazure}, {Meneux}, {Merighi},
  {Paltani}, {Pello}, {Pollo}, {Radovich}, {Bondi}, {Bongiorno}, {Busarello},
  {Ciliegi}, {Lamareille}, {Mellier}, {Merluzzi}, {Ripepi}, \&
  {Rizzo}}]{vvds_lumfunc}
{Ilbert}, O., {Tresse}, L., {Zucca}, E., {et~al.} 2005, \aap, 439, 863

\bibitem[{{Kauffmann} {et~al.}(2003{\natexlab{a}}){Kauffmann}, {Heckman},
  {White}, {Charlot}, {Tremonti}, {Brinchmann}, {Bruzual}, {Peng}, {Seibert},
  {Bernardi}, {Blanton}, {Brinkmann}, {Castander}, {Cs{\' a}bai}, {Fukugita},
  {Ivezic}, {Munn}, {Nichol}, {Padmanabhan}, {Thakar}, {Weinberg}, \&
  {York}}]{ste_mass_1}
{Kauffmann}, G., {Heckman}, T.~M., {White}, S.~D.~M., {et~al.}
  2003{\natexlab{a}}, \mnras, 341, 33

\bibitem[{{Kauffmann} {et~al.}(2003{\natexlab{b}}){Kauffmann}, {Heckman},
  {White}, {Charlot}, {Tremonti}, {Peng}, {Seibert}, {Brinkmann}, {Nichol},
  {SubbaRao}, \& {York}}]{bimodal_4000}
{Kauffmann}, G., {Heckman}, T.~M., {White}, S.~D.~M., {et~al.}
  2003{\natexlab{b}}, \mnras, 341, 54

\bibitem[{{Kaviraj} {et~al.}(2006){Kaviraj}, {Devriendt}, {Ferreras}, {Yi}, \&
  {Silk}}]{early_progenitors}
{Kaviraj}, S., {Devriendt}, E.~G., {Ferreras}, I., {Yi}, S.~K., \& {Silk}, J.
  2006, astro-ph/0602347

\bibitem[{{Kennicutt}(1998)}]{kenni_sfr}
{Kennicutt}, R.~C. 1998, \araa, 36, 189

\bibitem[{{Kodama} \& {Arimoto}(1997)}]{kodama_models}
{Kodama}, T. \& {Arimoto}, N. 1997, \aap, 320, 41

\bibitem[{{Kodama} {et~al.}(1999){Kodama}, {Bower}, \& {Bell}}]{redpop_hdf}
{Kodama}, T., {Bower}, R.~G., \& {Bell}, E.~F. 1999, \mnras, 306, 561

\bibitem[{{Le F{\` e}vre} {et~al.}(2004){Le F{\` e}vre}, {Mellier},
  {McCracken}, {Foucaud}, {Gwyn}, {Radovich}, {Dantel-Fort}, {Bertin},
  {Moreau}, {Cuillandre}, {Pierre}, {Le Brun}, {Mazure}, \&
  {Tresse}}]{vvds_imaging}
{Le F{\` e}vre}, O., {Mellier}, Y., {McCracken}, H.~J., {et~al.} 2004, \aap,
  417, 839

\bibitem[{{Le F{\`e}vre} {et~al.}(2005){Le F{\`e}vre}, {Vettolani}, {Garilli},
  {Tresse}, {Bottini}, {Le Brun}, {Maccagni}, {Picat}, {Scaramella},
  {Scodeggio}, {Zanichelli}, {Adami}, {Arnaboldi}, {Arnouts}, {Bardelli},
  {Bolzonella}, {Cappi}, {Charlot}, {Ciliegi}, {Contini}, {Foucaud},
  {Franzetti}, {Gavignaud}, {Guzzo}, {Ilbert}, {Iovino}, {McCracken}, {Marano},
  {Marinoni}, {Mathez}, {Mazure}, {Meneux}, {Merighi}, {Paltani}, {Pell{\`o}},
  {Pollo}, {Pozzetti}, {Radovich}, {Zamorani}, {Zucca}, {Bondi}, {Bongiorno},
  {Busarello}, {Lamareille}, {Mellier}, {Merluzzi}, {Ripepi}, \&
  {Rizzo}}]{vvds_main}
{Le F{\`e}vre}, O., {Vettolani}, G., {Garilli}, B., {et~al.} 2005, \aap, 439,
  845

\bibitem[{{LeFevre} {et~al.}(2003){LeFevre}, {Saisse}, {Mancini}, {Brau-Nogue},
  {Caputi}, {Castinel}, {D'Odorico}, {Garilli}, {Kissler-Patig}, {Lucuix},
  {Mancini}, {Pauget}, {Sciarretta}, {Scodeggio}, {Tresse}, \&
  {Vettolani}}]{vimos_comm}
{LeFevre}, O., {Saisse}, M., {Mancini}, D., {et~al.} 2003, in Instrument Design
  and Performance for Optical/Infrared Ground-based Telescopes. Edited by Iye,
  Masanori; Moorwood, Alan F. M. Proceedings of the SPIE, Volume 4841, pp.
  1670-1681 (2003)., ed. M.~{Iye} \& A.~F.~M. {Moorwood}, 1670--1681

\bibitem[{{Lilly} {et~al.}(1995){Lilly}, {Tresse}, {Hammer}, {Crampton}, \& {Le
  F{\` e}vre}}]{cfrs_iii}
{Lilly}, S.~J., {Tresse}, L., {Hammer}, F., {Crampton}, D., \& {Le F{\` e}vre},
  O. 1995, \apj, 455, 108

\bibitem[{{McCracken} {et~al.}(2003){McCracken}, {Radovich}, {Bertin},
  {Mellier}, {Dantel-Fort}, {Le F{\` e}vre}, {Cuillandre}, {Gwyn}, {Foucaud},
  \& {Zamorani}}]{vvds_imaging_f02}
{McCracken}, H.~J., {Radovich}, M., {Bertin}, E., {et~al.} 2003, \aap, 410, 17

\bibitem[{{Menci} {et~al.}(2005){Menci}, {Fontana}, {Giallongo}, \&
  {Salimbeni}}]{bimo_models}
{Menci}, N., {Fontana}, A., {Giallongo}, E., \& {Salimbeni}, S. 2005, \apj,
  632, 49

\bibitem[{{Meneux} {et~al.}(2006){Meneux}, {Le F{\`e}vre}, {Guzzo}, {Pollo},
  {Cappi}, {Ilbert}, {Iovino}, {Marinoni}, {McCracken}, {Bottini}, {Garilli},
  {Le Brun}, {Maccagni}, {Picat}, {Scaramella}, {Scodeggio}, {Tresse},
  {Vettolani}, {Zanichelli}, {Adami}, {Arnouts}, {Arnaboldi}, {Bardelli},
  {Bolzonella}, {Charlot}, {Ciliegi}, {Contini}, {Foucaud}, {Franzetti},
  {Gavignaud}, {Marano}, {Mazure}, {Merighi}, {Paltani}, {Pell{\`o}},
  {Pozzetti}, {Radovich}, {Zamorani}, {Zucca}, {Bondi}, {Bongiorno},
  {Busarello}, {Cucciati}, {Gregorini}, {Lamareille}, {Mathez}, {Mellier},
  {Merluzzi}, {Ripepi}, \& {Rizzo}}]{vvds_clustering}
{Meneux}, B., {Le F{\`e}vre}, O., {Guzzo}, L., {et~al.} 2006, \aap, 452, 387

\bibitem[{{Mignoli} {et~al.}(2005){Mignoli}, {Cimatti}, {Zamorani}, {Pozzetti},
  {Daddi}, {Renzini}, {Broadhurst}, {Cristiani}, {D'Odorico}, {Fontana},
  {Giallongo}, {Gilmozzi}, {Menci}, \& {Saracco}}]{k20_spectro}
{Mignoli}, M., {Cimatti}, A., {Zamorani}, G., {et~al.} 2005, \aap, 437, 883

\bibitem[{{Radovich} {et~al.}(2004){Radovich}, {Arnaboldi}, {Ripepi},
  {Massarotti}, {McCracken}, {Mellier}, {Bertin}, {Zamorani}, {Adami},
  {Bardelli}, {Le F{\` e}vre}, {Foucaud}, {Garilli}, {Scaramella}, {Vettolani},
  {Zanichelli}, \& {Zucca}}]{vvds_imaging_u}
{Radovich}, M., {Arnaboldi}, M., {Ripepi}, V., {et~al.} 2004, \aap, 417, 51

\bibitem[{{Renzini}(2006)}]{renzini_review}
{Renzini}, A. 2006, \araa, 44, 141

\bibitem[{{Roberts} \& {Haynes}(1994)}]{color_morph}
{Roberts}, M.~S. \& {Haynes}, M.~P. 1994, \araa, 32, 115

\bibitem[{{Salpeter}(1955)}]{salpeter}
{Salpeter}, E.~E. 1955, \apj, 121, 161

\bibitem[{{Sandage}(1986)}]{sandage}
{Sandage}, A. 1986, \aap, 161, 89

\bibitem[{{Schlegel} {et~al.}(1998){Schlegel}, {Finkbeiner}, \&
  {Davis}}]{dust_map}
{Schlegel}, D.~J., {Finkbeiner}, D.~P., \& {Davis}, M. 1998, \apj, 500, 525

\bibitem[{{Scodeggio} {et~al.}(2005){Scodeggio}, {Franzetti}, {Garilli},
  {Zanichelli}, {Paltani}, {Maccagni}, {Bottini}, {Le Brun}, {Contini},
  {Scaramella}, {Adami}, {Bardelli}, {Zucca}, {Tresse}, {Ilbert}, {Foucaud},
  {Iovino}, {Merighi}, {Zamorani}, {Gavignaud}, {Rizzo}, {McCracken}, {Le
  F{\`e}vre}, {Picat}, {Vettolani}, {Arnaboldi}, {Arnouts}, {Bolzonella},
  {Cappi}, {Charlot}, {Ciliegi}, {Guzzo}, {Marano}, {Marinoni}, {Mathez},
  {Mazure}, {Meneux}, {Pell{\`o}}, {Pollo}, {Pozzetti}, \& {Radovich}}]{vipgi}
{Scodeggio}, M., {Franzetti}, P., {Garilli}, B., {et~al.} 2005, \pasp, 117,
  1284

\bibitem[{{Stanford} {et~al.}(1998){Stanford}, {Eisenhardt}, \&
  {Dickinson}}]{mono_test_4}
{Stanford}, S.~A., {Eisenhardt}, P.~R., \& {Dickinson}, M. 1998, \apj, 492, 461

\bibitem[{{Strateva} {et~al.}(2001){Strateva}, {Ivezi{\' c}}, {Knapp},
  {Narayanan}, {Strauss}, {Gunn}, {Lupton}, {Schlegel}, {Bahcall}, {Brinkmann},
  {Brunner}, {Budav{\' a}ri}, {Csabai}, {Castander}, {Doi}, {Fukugita}, {Gy{\H
  o}ry}, {Hamabe}, {Hennessy}, {Ichikawa}, {Kunszt}, {Lamb}, {McKay},
  {Okamura}, {Racusin}, {Sekiguchi}, {Schneider}, {Shimasaku}, \&
  {York}}]{bimodal_1}
{Strateva}, I., {Ivezi{\' c}}, {\v Z}., {Knapp}, G.~R., {et~al.} 2001, \aj,
  122, 1861

\bibitem[{{Tremonti} {et~al.}(2004){Tremonti}, {Heckman}, {Kauffmann},
  {Brinchmann}, {Charlot}, {White}, {Seibert}, {Peng}, {Schlegel}, {Uomoto},
  {Fukugita}, \& {Brinkmann}}]{mass_metallicity}
{Tremonti}, C.~A., {Heckman}, T.~M., {Kauffmann}, G., {et~al.} 2004, \apj, 613,
  898

\bibitem[{{Tully} {et~al.}(1982){Tully}, {Mould}, \& {Aaronson}}]{color_mag2}
{Tully}, R.~B., {Mould}, J.~R., \& {Aaronson}, M. 1982, \apj, 257, 527

\bibitem[{{van Dokkum} \& {Franx}(2001)}]{progenitor_bias}
{van Dokkum}, P.~G. \& {Franx}, M. 2001, \apj, 553, 90

\bibitem[{{Visvanathan} \& {Sandage}(1977)}]{color_mag1}
{Visvanathan}, N. \& {Sandage}, A. 1977, \apj, 216, 214

\bibitem[{{Weiner} {et~al.}(2005){Weiner}, {Phillips}, {Faber}, {Willmer},
  {Vogt}, {Simard}, {Gebhardt}, {Im}, {Koo}, {Sarajedini}, {Wu}, {Forbes},
  {Gronwall}, {Groth}, {Illingworth}, {Kron}, {Rhodes}, {Szalay}, \&
  {Takamiya}}]{bimodal_6}
{Weiner}, B.~J., {Phillips}, A.~C., {Faber}, S.~M., {et~al.} 2005, \apj, 620,
  595

\bibitem[{{Zanichelli} {et~al.}(2005){Zanichelli}, {Garilli}, {Scodeggio},
  {Franzetti}, {Rizzo}, {Maccagni}, {Merighi}, {Picat}, {Le F{\`e}vre},
  {Foucaud}, {Bottini}, {Le Brun}, {Scaramella}, {Tresse}, {Vettolani},
  {Adami}, {Arnaboldi}, {Arnouts}, {Bardelli}, {Bolzonella}, {Cappi},
  {Charlot}, {Ciliegi}, {Contini}, {Gavignaud}, {Guzzo}, {Ilbert}, {Iovino},
  {McCracken}, {Marano}, {Marinoni}, {Mathez}, {Mazure}, {Meneux}, {Paltani},
  {Pell{\`o}}, {Pollo}, {Pozzetti}, {Radovich}, {Zamorani}, \& {Zucca}}]{ifu}
{Zanichelli}, A., {Garilli}, B., {Scodeggio}, M., {et~al.} 2005, \pasp, 117,
  1271

\bibitem[{{Zucca} {et~al.}(2006){Zucca}, {Ilbert}, {Bardelli}, {Tresse},
  {Zamorani}, {Arnouts}, {Pozzetti}, {Bolzonella}, {McCracken}, {Bottini},
  {Garilli}, {Le Brun}, {Le F{\`e}vre}, {Maccagni}, {Picat}, {Scaramella},
  {Scodeggio}, {Vettolani}, {Zanichelli}, {Adami}, {Arnaboldi}, {Cappi},
  {Charlot}, {Ciliegi}, {Contini}, {Foucaud}, {Franzetti}, {Gavignaud},
  {Guzzo}, {Iovino}, {Marano}, {Marinoni}, {Mazure}, {Meneux}, {Merighi},
  {Paltani}, {Pell{\`o}}, {Pollo}, {Radovich}, {Bondi}, {Bongiorno},
  {Busarello}, {Cucciati}, {Gregorini}, {Lamareille}, {Mathez}, {Mellier},
  {Merluzzi}, {Ripepi}, \& {Rizzo}}]{lum_type_vvds}
{Zucca}, E., {Ilbert}, O., {Bardelli}, S., {et~al.} 2006, \aap, 455, 879

\end{thebibliography}
%\end{thebibliography}

\end{document}